\documentstyle[epsf,aps,german]{revtex}

\dateenglish
\captionsenglish
\def\la{\mathrel{\mathchoice{\vcenter{\offinterlineskip\halign{\hfil
$\displaystyle##$\hfil\cr<\cr\sim\cr}}}
{\vcenter{\offinterlineskip\halign{\hfil$\textstyle##$\hfil\cr<\cr\sim\cr}}}
{\vcenter{\offinterlineskip\halign{\hfil$\scriptstyle##$\hfil\cr<\cr\sim\cr}}}
{\vcenter{\offinterlineskip\halign{\hfil$\scriptscriptstyle##$\hfil\cr<\cr\sim
\cr}}}}}
\def\ga{\mathrel{\mathchoice{\vcenter{\offinterlineskip\halign{\hfil
$\displaystyle##$\hfil\cr>\cr\sim\cr}}}
{\vcenter{\offinterlineskip\halign{\hfil$\textstyle##$\hfil\cr>\cr\sim\cr}}}
{\vcenter{\offinterlineskip\halign{\hfil$\scriptstyle##$\hfil\cr>\cr\sim\cr}}}
{\vcenter{\offinterlineskip\halign{\hfil$\scriptscriptstyle##$\hfil\cr>\cr\sim
\cr}}}}}
\def\laq{\mathrel{\mathchoice{\vcenter{\offinterlineskip\halign{\hfil
$\displaystyle##$\hfil\cr<\cr=\cr}}}
{\vcenter{\offinterlineskip\openup
0,5mm\halign{\hfil$\textstyle##$\hfil\cr<\cr=\cr}}}
{\vcenter{\offinterlineskip\halign{\hfil$\scriptstyle##$\hfil\cr<\cr=\cr}}}
{\vcenter{\offinterlineskip\halign{\hfil$\scriptscriptstyle##$\hfil\cr<\cr=
\cr}}}}} 
\def\gaq{\mathrel{\mathchoice{\vcenter{\offinterlineskip\halign{\hfil
$\displaystyle##$\hfil\cr>\cr=\cr}}}
{\vcenter{\offinterlineskip\openup
0,5mm\halign{\hfil$\textstyle##$\hfil\cr>\cr=\cr}}}
{\vcenter{\offinterlineskip\halign{\hfil$\scriptstyle##$\hfil\cr>\cr=\cr}}}
{\vcenter{\offinterlineskip\halign{\hfil$\scriptscriptstyle##$\hfil\cr>\cr=
\cr}}}}}

\draft

\begin{document}
\title{\vspace*{-2.0cm}\hfill {\tt LPTHE Orsay 93/72\\ }
  {}\hfill {\tt cond-mat/9507092}
  \vspace*{0.5cm}
  \\
Theory of the critical state of low-dimensional spin glass}

\author{M.J.~Thill and H.J.~Hilhorst}

\address{Laboratoire de Physique Th\'eorique et Hautes Energies
(CNRS - URA 63),\\ B\^atiment 211, Universit\'e de Paris-Sud, F-91405 Orsay 
CEDEx}

\address{\em (\today)}

\address{
\centering{
\medskip\em
\begin{minipage}{14cm}
We analyse the critical region of finite-($d$)-dimensional Ising spin glass, 
in particular the limit of $d$ closely above the lower critical dimension $d_
\ell$. At criticality the thermally active degrees of freedom are surfaces
(of width zero) enclosing clusters of spins that may reverse with respect
to their environment. The surfaces are organised in finite interacting
structures. These may be called {\em protodroplets}\/, since in the
off-critical limit they reduce to the Fisher and Huse droplets. This 
picture provides an explanation for the phenomenon of critical chaos 
discovered earlier. It also implies that the spin-spin and energy-energy 
correlation functions are multifractal and we present scaling laws that
describe them. Several of our results should be verifiable 
in Monte Carlo studies at finite temperature in $d=3$.
\pacs{\noindent PACS numbers: 05.50.+q, 64.60.Fr, 75.50.Lk}
\end{minipage}
}}

\maketitle

\section{Introduction}

No exact solution exists of the spin glass problem in the physical
dimensions $d=2$ and $d=3$. The most successful approximate description
was developed by McMillan \cite{MM}, Bray and Moore \cite{BM}, and Fisher
and Huse \cite{FH,FHua}. The authors \cite{BM} speak of a {\em scaling}\/ 
theory and the authors \cite{FH,FHua} of a {\em droplet}\/ theory. 
In both cases the main hypothesis is the existence of 
an effective coupling constant $K^L$ dependent on the length scale $L$ and 
satisfying scaling laws near zero temperature (with thermal exponent $y$) 
and at criticality (with exponent $y_c$). These scaling laws 
may also be thought of as arising from a renormalisation group (RG) for the 
spin glass, and in fact real space RG arguments are used by all 
the authors \cite{MM,BM,FH,FHua}. In particular, RG arguments can be put 
forward \cite{BM86a} that experimental spin glass is well described by the 
Edwards-Anderson (EA) {\em Ising}\/ spin glass model \cite{EA}.

Again, no exact RG is known for the finite-dimensional spin glass
in dimensions $d>1$. The most popular approximate RG is the Migdal-Kadanoff 
(MK) construction \cite{MK,YS,SY,AP}. 
In spite of its evident shortcomings the MKRG has been successful 
as an approximate calculation scheme for ferromagnets and other 
nonrandom systems. For the finite-dimensional spin glass there 
does not today exist a better alternative. 
 
Upon comparison one finds that the MKRG results for the spin 
glass are close, if not identical, to those obtained by
the scaling and droplet theory. Most conclusions from MKRG are easily 
interpreted in the language of the droplet theory and the differences where
they appear can be easily understood. There is therefore no reason
at this level of approximation not to work with the MKRG,
or, for that matter, with any real space RG of the same ilk.
In this work we shall exploit the potential of the MKRG for spin glasses
more fully than has been done hitherto. Our aim is to explain the phenomenon
of ``chaos" known to occur at criticality in low enough dimensions 
\cite{NH1,NH2}. We achieve this goal by our interpretation of new MKRG 
results.
We are led, in particular, to a characterisation of the thermally active 
degrees of freedom in critical Ising spin glass just above their lower 
critical
dimension.

In order to set the stage for our work, we recall the most essential ones of 
the established results on zero temperature and critical chaos. 

A {\em droplet}\/ of linear size $L$ at a point $\mbox{\boldmath $R$}$ 
in space is defined as the domain of $\sim L^d$ spins around this point that 
can be reversed with respect to its environment at minimum free energy cost.
The droplets are the low free energy excitations upon which the theory
by Fisher and Huse \cite{FH,FHua} is based. Droplet walls have an 
intrinsic width of the order of the correlation length $\xi_-$ and hence 
the droplet concept is naturally limited to length scales $L\ga\xi_-(T)$. In 
particular, it no longer applies in the limit $T\nearrow T_c$, where $
\xi_-(T)$ diverges.

Spin glasses have the property of being chaotic, as was revealed in 
the work of Bray and Moore \cite{BM}, Fisher and Huse \cite{FH}, and 
Banavar and Bray \cite{BB}. The term {\em chaotic}\/ here denotes the 
property 
that the thermally averaged relative orientation of two spins a distance
$L$ apart is a rapidly and randomly varying function of temperature (see
figure~1).
If we denote the autocorrelation interval on the temperature axis of 
this function by $\Delta T^L$, then in the ordered phase
\begin{equation}\label{lTchaos}
\Delta T^L \sim  L^{-\zeta}\,\, ,\qquad  \mbox{as}\,\,  L\to\infty\,\, ,
\end{equation}
where $\zeta$ is the {\em chaos exponent}\/. Upon inverting (\ref{lTchaos}) 
one finds
\begin{equation}\label{ollT}
L_{\Delta T} \sim \Delta T^{-\frac{1}{\zeta}}\,\, ,\qquad  \mbox{as}\,\,  
\Delta T\to 0\,\, ,
\end{equation}
which is called the {\em overlap length}\/ associated with a temperature 
difference $\Delta T$. Numerically $\zeta$ takes values between $0.75$ and
$1$ in $1\laq d\laq 3$ \cite{BMBC,BMyc,CBC,NH1,NH2}. Within the
droplet theory the chaos phenomenon is elegantly explained
as resulting from temperature changes that disturb the energy versus entropy 
balance of the droplet walls and thereby induce random droplet reversals. In 
this theory the sum $2y+2\zeta$ of the zero temperature thermal and chaos
exponents appears to equal the fractal surface dimension $d_s$ of the droplet
walls.

The original authors \cite{MM,BM,FH} thought that chaos was limited to 
length scales $L\ga\xi_-$ in the spin glass ordered phase and to dimensions 
where such a phase exists, i.e., above the lower critical dimension $d_
\ell$. Ney-Nifle and Hilhorst \cite{NH1,NH2} subsequently showed that there 
is chaos also in the critical region, at least for spatial dimensions $d$ in 
an interval
extending upward from the lower critical dimension $d_\ell\approx 2.5$ 
\cite{SY,BM,SF} to some limit dimension $d_+$, 
estimated within MKRG to be $d_+\approx 3.4$ \cite{NH1,NH2}.

The reasoning can be summarised as follows. At a length scale $L$, 
the critical region is the temperature interval
\begin{equation}\label{yc}
\frac{|T-T_c|}{T_c} \la L^{-y_c}
\end{equation}
in which $L\la\xi_-(T)$.
Here $1/y_c\equiv\nu$ is the correlation length exponent. 
Ney-Nifle and Hilhorst use MKRG to calculate the effective couplings on 
scale $L$ and find that inside the critical region they
vary randomly with temperature with an autocorrelation interval 
\begin{equation}\label{zetac}
\Delta T^L \sim L^{-\zeta_c}\,\, ,
\end{equation}
where the (positive) {\em critical point chaos exponent}\/ $\zeta_c$ is a
new and independent exponent (see figure~1). This chaotic behaviour of the 
renormalised coupling $K^L$ ({\em chaos with temperature}\/) is reflected in 
the behaviour of the two-point correlation
function $<s_i s_{i+L}>$ as a function of temperature, which changes sign
when the effective coupling at distance $L$ changes sign. 

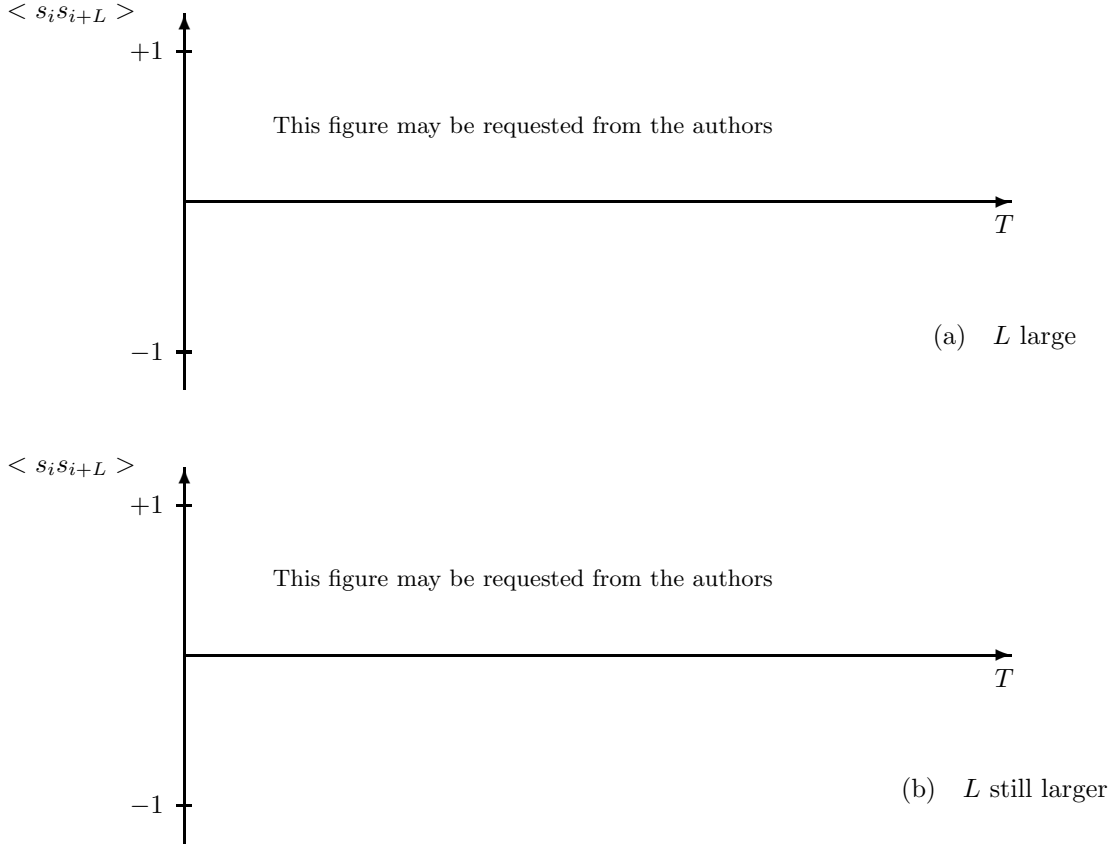
\begin{figure}[htb]
\setlength{\unitlength}{1cm}
\centering{
\begin{picture}(11,5)
\thicklines
\put(0.5,0){\vector(0,1){5}}
\put(0.5,2.5){\vector(1,0){11}}
\put(-1,5){\makebox(0,0){$<s_i s_{i+L}>$}}
\put(11.4,2.2){\makebox(0,0){$T$}}
\put(11.4,0.7){\makebox(0,0){(a)\quad $L$ large}}
\put(0.4,0.5){\line(1,0){0.2}}
\put(0.4,4.5){\line(1,0){0.2}}
\put(0,0.5){\makebox(0,0){$-1$}}
\put(0,4.5){\makebox(0,0){$+1$}}
\put(5,3.5){\makebox(0,0){\small This figure may be requested from the
authors}}
\end{picture}
}
\vspace{1cm}
\centering{
\begin{picture}(11,5)
\thicklines
\put(0.5,0){\vector(0,1){5}}
\put(0.5,2.5){\vector(1,0){11}}
\put(-1,5){\makebox(0,0){$<s_i s_{i+L}>$}}
\put(11.4,2.2){\makebox(0,0){$T$}}
\put(11.4,0.7){\makebox(0,0){(b)\quad $L$ still larger}}
\put(0.4,0.5){\line(1,0){0.2}}
\put(0.4,4.5){\line(1,0){0.2}}
\put(0,0.5){\makebox(0,0){$-1$}}
\put(0,4.5){\makebox(0,0){$+1$}}
\put(5,3.5){\makebox(0,0){\small This figure may be requested from the
authors}}
\end{picture}
}
\vspace{5mm}

\caption{Temperature dependence of the two-point correlation function
$<s_i s_{i+L}>$ for (a) large $L$ and (b) still larger $L$. At criticality $
\Delta T^L \sim L^{-\zeta_c}$ and near zero temperature $\Delta T^L \sim L^{-
\zeta}$, for $L\to \infty$.
}
\end{figure}

In view of this
there is chaos inside the critical region if 
\begin{equation}
\zeta_c > y_c \,\, .
\end{equation}
Figure~2 shows the dimension dependence of $y_c$ and $\zeta_c$. As $d
\searrow d_\ell$, where $d_\ell$ is characterised by $y(d_\ell)=0$, these 
exponents join continuously their zero temperature equivalents $y(d)$ and $
\zeta(d)$. The MKRG values for the
critical exponents in dimension $d=3$ are \cite{NH1,NH2,TH}
\begin{equation}
y_c(3)\approx 0.36, \quad \zeta_c(3)\approx 0.57\,\, .
\end{equation}
Hence, according to these values, the $3d$ spin glass is chaotic at 
criticality.
\newpage
\begin{figure}[htb]
\vspace{0.5cm}
\parbox[b]{15.5cm}{${}$\hspace{6.9cm} $d_\ell$\hspace{3.1cm} $d_+$}
\vspace{3.8cm}

\parbox[b]{15.5cm}{${}$\hspace{5.2cm} $2$\hspace{3.5cm} $3$}
\vspace{-8.8cm}

\begin{minipage}[h]{15.5cm}
\begin{minipage}[t]{2cm}
\makebox[0cm]{}
${}$\\[5.3cm]
\begin{tabular}{r}
${}$\hspace{1cm}$0.5$\\[3.4cm]
$-0.5$
\end{tabular}
\end{minipage}\hspace{-0.9cm}
\begin{minipage}[t]{11cm}
\makebox[0cm]{}
  \epsfxsize=11cm
	\epsfbox{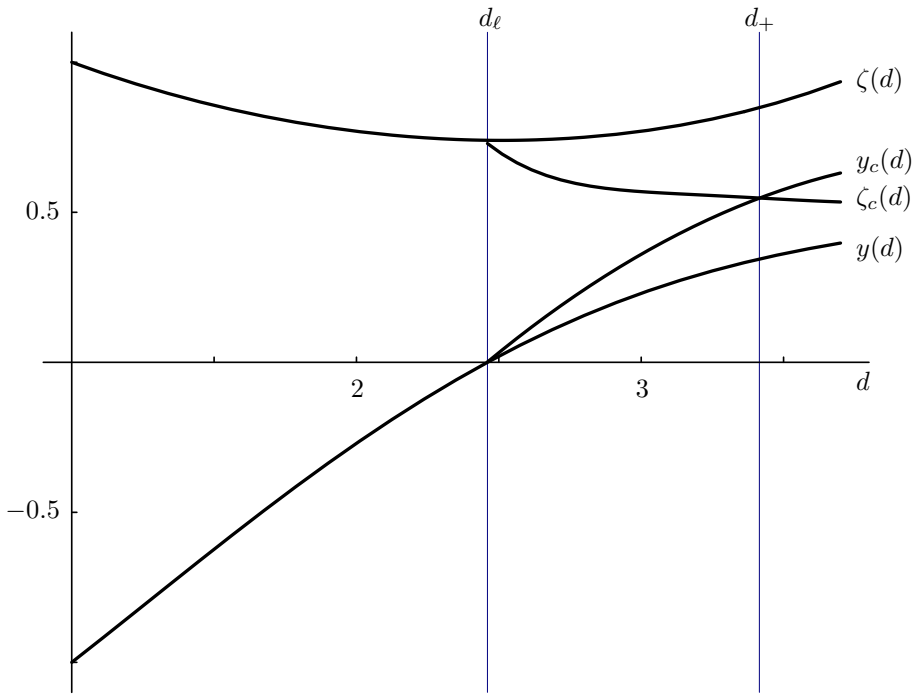}
\end{minipage}\hspace{-0.3cm}
\begin{minipage}[t]{1cm}
\makebox[0cm]{}
${}$\\[3.5cm]
$\zeta(d)$\\[0.6cm]
$y_c(d)$\\
$\zeta_c(d)$\\[0.2cm]
$y(d)$\\[1.3cm]
$d$
\end{minipage}\hfill
\begin{minipage}[t]{2cm}
\makebox[0cm]{}
\end{minipage}
\end{minipage}
\vspace{-2.5cm}

\caption{The critical exponents $y_c(d)$ and $\zeta_c(d)$ as a function of 
the 
dimension $d$. They join continuously the zero temperature exponents
$y(d)$ and $\zeta(d)$, respectively, as $d$ approaches
$d_\ell$ from above. This dimension is defined by $y(d_\ell)=0$. The 
intersection of the curves $y_c(d)$ and $\zeta_c(d)$ defines $d_+$. The 
curves shown are obtained from numerical evaluation of MKRG by the
pool method \protect\cite{TH}.}
\end{figure}

The problem that triggered the investigation of this paper can now be stated 
precisely. In view of the inapplicability of the droplet concept in the 
critical region,
the natural question to ask is if the chaos at criticality can be given some 
different geometric 
interpretation. Blind analogy might lead one to suppose, for example,
that the sum $2y_c+2\zeta_c$ would be equal to the fractal surface
dimension of objects yet to be identified whose reversal upon 
temperature changes would cause the critical chaos - however we
shall see that this is not true. This work deals with answering the question 
raised above, but its results go beyond it:

The paper is organised as follows. In section~\ref{gen} we review and extend 
the existing MKRG theory of the Ising 
spin glass. Introducing the scaling of various averaged derivatives 
of the renormalised coupling constant $K^L$, we obtain multifractal 
scaling of the energy-energy correlations at criticality. 
In section~\ref{pi} we show how the scaling laws
deduced from MKRG can be given a geometric interpretation. The fundamental
entities are surfaces (of width zero) enclosing clusters of spins that may
reverse with respect to their environment. The surfaces are grouped into
finite interacting sets that are self-similar in the range of scales between
the lattice cut-off and their own size. Such a set of interacting surfaces
is appropriately called a {\em protodroplet}\/, since in the off-critical
limit it reduces to a Fisher and Huse droplet. Near the lower critical 
dimension there is confluency of zero
temperature and critical phenomena. In order to clarify the picture, the 
characteristic lengths that play a r\^ole are analysed in section~\ref{char}.
In section~\ref{ssmf}, based on the physical picture of section~\ref{pi}, 
we establish the multifractal scaling of the 
spin-spin correlation function at criticality. Furthermore, 
the effects of a magnetic field in the ordered phase 
and at criticality are discussed in a unified way, within the 
picture of section~\ref{gen}. We use an Imry-Ma type argument to 
deduce scaling laws for the leading behaviour of the magnetisation in 
the neighbourhood of the critical point, approaching $T_c$ from below 
and above. The singular situation at $d_\ell$ is also considered in 
detail. In section~\ref{dyn} we point out the consequences of the 
strong spatial inhomogeneity of thermal fluctua\-ti\-ons on the critical
dynamics. The picture of this paper augmented with the additional
assumption that the two-dimensional dynamics near zero temperature
dynamics is activated, together with an argument of
continuity, gives a first strong theoretical support for the conjecture 
that the critical dynamics of $d$-dimensional spin glass is activated 
in an interval of dimensions starting with $d_\ell$. In the conclusion, 
section~\ref{concl}, we mention possibilties of verifying our picture
in simulations of the $d=3$ EA model.

\setcounter{equation}{0}
\section{Extended scaling laws for renormalised couplings}\label{gen}

\subsection{Introduction}

The Migdal-Kadanoff renormalisation transformation \cite{MK} was first 
formulated for the spin glass by Young {\sl et al}\/ \cite{YS,SY}. In this 
section we start from the same basic formulas. We briefly recall the main 
results known and then go on to
extend some of them. Our interest concerns in particular correlations 
of various finds (with respect to the disorder average) between the 
renormalised and the unrenormalised couplings. 

The MKRG works as follows. For a rescaling factor $b$ one combines a set of 
$b^d$ couplings at the $n$th level, $\{K_{i_n}^{(n)}\}_{i_{n+1}}$, to a new 
coupling, $K_{i_{n+1}}^{(n+1)}$. This is done by arranging the 
$K_{i_n}^{(n)}$ of the block $i_{n+1}$ in $b^{d-1}$ parallel strings 
(labeled by the index $m$) of $b$ couplings each. [Whenever we will indicate 
a RG step explicitly in thwhat follows we will relabel the couplings of $
\{K_{i_n}^{(n)}\}_{i_{n+1}}$ as $K_{mk}^{(n)}$.] The recursion relation then 
reads \cite{SY,CE}
\begin{equation}\label{renK}
K_{i_{n+1}}^{(n+1)} = \sum\limits_{m=1}^{b^{d-1}} \mbox{arctanh}\prod
\limits_{k=1}^b \tanh K_{mk}^{(n)}\,\, .
\end{equation}
In the remainder we set $b=2$. The couplings $K_{i_0}^{(0)}$ of the 
originally given problem are of the form $K_i\equiv\frac{J_i}{k_{\mbox{\tiny 
B}}T}$, where $i$ is a bond index, and the $J_i$ are the coupling constants 
on the original lattice with root-mean-square (rms) value $J$. We shall also 
use $K\equiv{\overline{K_i^2}}^\frac{1}{2}\equiv\beta J$ as the temperature 
variable. With the $J_i$ given, the transformation (\ref{renK}) determines 
uniquely the couplings at all higher levels $n=1,2,\ldots$. For the $n$ 
times renormalised value of $K$ we shall also write
\begin{equation}
K^L \equiv K^{(n)} \quad \mbox{when} \quad L = 2^n 
\end{equation}
and for a specific scale-$L$ coupling\marginpar{disc}
\begin{equation}
K^L_{\mbox{\boldmath\tiny $R$}} \equiv K_{i_n}^{(n)} 
\end{equation}
when $i_n$ is the label of the coupling at $\mbox{\boldmath $R$}$.
$K^{(n)}$ can be calculated from the unrenormalised rms coupling 
$K \equiv K^{(0)}$ via the transformation (\ref{renK}). 
If, as usual, after each RG 
step the distribution of couplings is replaced by a Gaussian with the 
same variance, one finds the recursion
\begin{equation}
\begin{array}{lcl}
K^{(n+1)} &=& 2^\frac{d-1}{2}\, \left[\,\,\overline{\mbox{arctanh}^2\prod
\limits_{k=1}^2 \tanh K_{mk}^{(n)}}\,\,\right]^\frac{1}{2}\\[4mm]
{} &\equiv& {\cal R}(K^{(n)})\,\, .
\end{array}
\end{equation}

The transformation ${\cal R}$ has the following properties (for some 
mathematically rigorous proofs see \cite{CE} and references therein). It has 
the fixed points $K=0$ and $K=\infty$, and, for dimensions $d$ above $d_
\ell$, also a critical fixed point $K=K_c$. Near the zero-$T$ fixed point 
one has  
\begin{equation}\label{Ky}
K^{(n+1)} \simeq 2^{y}K^{(n)}\,\, ,
\end{equation}
with $y < 0$ for $d < d_\ell$ and $y > 0$ for $d > d_\ell$, and near the 
critical fixed point $K=K_c$ 
\begin{equation}\label{RGTc}
K^{(n+1)} - K_c \simeq 2^{y_c}(K^{(n)} - K_c)\,\, . 
\end{equation}
By expanding (\ref{renK}) for $\beta\to\infty$, one easily shows that 
closely above the lower critical dimension the equations (\ref{Ky}) and (
\ref{RGTc}) can be written together in the compact form 
\begin{equation}\label{renKlT}
(K^L)^2 = 2^{2y} (K^\frac{L}{2})^2 - c_0
\end{equation}
where $c_0$ is a positive constant.

\subsection{Multifractal scaling at criticality}

Let $K^L_{\mbox{\boldmath\tiny $R$}}$ be an individual renormalised 
coupling on scale $L$. One main result of reference
\cite{NH1,NH2} is that in the critical region 
$K^L_{\mbox{\boldmath\tiny $R$}}$ is a rapidly and randomly varying 
function of temperature with rms derivative
\begin{equation}\label{Ezetac}
\overline{\left(\frac{dK^L_{\mbox{\boldmath\tiny $R$}}}{dK}\right)^2} \sim 
L^{2\zeta_c}\,\, ,\quad\qquad
\left|\frac{K-K_c}{K_c}\right| \la L^{-y_c}
\end{equation}
in an interval of dimensions $d_\ell< d < d_+$. 
This chaotic behaviour of the renormalised coupling $K^L_{\mbox{\boldmath
\tiny $R$}}$ is reflected in the behaviour of the two-point correlation
function $<s_i s_{i+L}>$ as a function of temperature, which changes sign
when the effective coupling at distance $L$ changes sign (see figure~1). 
From (\ref{RGTc}) we have 
\begin{equation}\label{Eyc}
\frac{d}{dK}\overline{(K^L_{\mbox{\boldmath\tiny $R$}})^2} \sim L^{y_c}\,\, .
\end{equation}
Upon rewriting (\ref{Eyc}) as 
\begin{equation}\label{Entyc}
\overline{K^L_{\mbox{\boldmath\tiny $R$}}\frac{dK^L_{\mbox{\boldmath\tiny 
$R$}}}{dK}} \sim L^{y_c}\,\, ,
\end{equation}
one sees that $K^L_{\mbox{\boldmath\tiny $R$}}$ and $\frac{dK^L_{\mbox{
\boldmath\tiny $R$}}}{dK}$ are correlated, while for
reasons of symmetry  
\begin{equation}
\overline{\frac{dK^L_{\mbox{\boldmath\tiny $R$}}}{dK}}=0\,\, .
\end{equation}
The relations (\ref{Ezetac}) and (\ref{Entyc}) imply that, at criticality $
\frac{dK^L_{\mbox{\boldmath\tiny $R$}}}{dK}$ has a large random part, 
proportional to $L^{\zeta_c}$, which has a random sign, and a small random 
part, proportional to $L^{y_c}$, whose sign is positively correlated with 
that
of $K^L_{\mbox{\boldmath\tiny $R$}}$ (and which is therefore, in this sense, 
`systematic').

Below we shall study the joint probability distribution at criticality of  
$K^L_{\mbox{\boldmath\tiny $R$}}$ and $\frac{\partial K^L_{\mbox{\boldmath
\tiny $R$}}}{\partial K_i}K_i$. We will argue that at criticality $\zeta_c$ 
is only the first of a family of an infinite number of exponents describing, 
together with $y_c$, the scaling of the moments of the joint distribution. 
This is in contrast to the behaviour at zero temperature whose scaling is 
entirely described by the exponents $y$ and $\zeta$. 

Since we wish to consider two-point correlation functions in section 
\ref{ef}, we investigate the dependence of the
renormalised coupling $K^L_{\mbox{\boldmath\tiny $R$}}$ at a position $\mbox{
\boldmath $R$}$ and at length scale $L$ on the set
of unrenormalised couplings $\{K_i\}$ in the volume of order $L^d$ around $
\mbox{\boldmath $R$}$. 
The temperature derivative of $K^L_{\mbox{\boldmath\tiny $R$}}$ can be 
calculated via (\ref{renK}) as  
\begin{equation}\label{td}
\frac{dK^L_{\mbox{\boldmath\tiny $R$}}}{dK} = \sum\limits_i \frac{\partial 
K^L_{\mbox{\boldmath\tiny $R$}}}{\partial
K_i}\frac{K_i}{K}
\end{equation}
where the sum is on the $L^d$ unrenormalised couplings. Neglecting cross 
terms (that contribute only to order $L^{2y_c}$ as $L\rightarrow\infty$), we 
find from (\ref{Ezetac}) and (\ref{td}) at criticality
\begin{equation}\label{m1}
\overline{\left(\frac{\partial K^L_{\mbox{\boldmath\tiny $R$}}}{\partial 
K_i}K_i\right)^2} \sim
L^{2\zeta_c-d}\, ,
\end{equation}
Similarly (\ref{Entyc}) and (\ref{td}) give
\begin{equation}\label{m2}
\overline{K^L_{\mbox{\boldmath\tiny $R$}}\frac{\partial K^L_{\mbox{\boldmath
\tiny $R$}}}{\partial K_i} K_i} \sim L^{y_c-d}\, .
\end{equation}

The relations (\ref{m1}) and (\ref{m2}) are only two moments of a joint
probability distribution of $K^L_{\mbox{\boldmath\tiny $R$}}$ and $\frac{
\partial K^L_{\mbox{\boldmath\tiny $R$}}}{\partial K_i}K_i$.
For the other moments at the critical point we make the ansatz
\begin{equation}\label{em}
\overline{\left(\frac{\partial K^L_{\mbox{\boldmath\tiny $R$}}}{\partial 
K_i}K_i\right)^{2q}} \sim L^{2\zeta_c^{(q)}-d}
\end{equation}
and
\begin{equation}\label{om}
\overline{K^L_{\mbox{\boldmath\tiny $R$}}\left(\frac{\partial K^L_{\mbox{
\boldmath\tiny $R$}}}{\partial K_i}K_i\right)^{2q-1}} \sim
L^{y_c+2\zeta_c^{(q)}-2\zeta_c-d}\,\, ,
\end{equation}
where $\zeta_c^{(1)}\equiv \zeta_c$. 

Let now $K^L_{\mbox{\boldmath\tiny $R$}}\equiv K_{i_n}^{(n)}$ and $K_i\equiv 
K_{i_0}^{(0)}$ and write by chain rule differentiation down the hierarchy of 
couplings
\begin{equation}\label{cascK}
\frac{\partial K^L_{\mbox{\boldmath\tiny $R$}}}{\partial K_i} = \prod
\limits_{\nu=1}^n\frac{\partial
K_{i_\nu}^{(\nu)}}{\partial K_{i_{\nu-1}}^{(\nu-1)}}
\end{equation}
where $i_1, i_2, \ldots$ denote the locations of the couplings on the
successive levels $\nu=1,2,\ldots$ to which $K_i$ contributes.
At criticality, the factors in the product are
identically distributed random variables. If we assume them
independent, then
we have from (\ref{em}) and (\ref{cascK}) that at criticality
\begin{equation}\label{defzetaq}
2\zeta_c^{(q)}-d = (\log
2)^{-1}\log\overline{\left(\frac{\partial K_{i_1}}{\partial
K_{i_0}}\right)^{2q}}\,\, .
\end{equation}
The right-hand-side of this expression can be calculated from a single 
renormalisation iteration after which $\zeta_c^{(q)}$ is known \cite{DL}. 
Obviously there is no reason for $\zeta_c^{(q)}$ to be linear in $q$. Hence 
we have found that at criticality there is {\em multifractal scaling}\/ 
\cite{DL}. 

The above results are valid at criticality. It is instructive to compare 
them to their analogues in the ordered phase. There we have for $L\gg \xi_-$ 
\cite{NH1,NH2,FH}:
\begin{equation}\label{zTsc}
\begin{array}{lcl}
\overline{\left(\frac{dK^L_{\mbox{\boldmath\tiny $R$}}}{dK}\right)^2} &\sim& 
L^{d_s}\,\, ,\\[2mm]
\frac{d}{dK}\overline{(K^L_{\mbox{\boldmath\tiny $R$}})^2} &\sim& L^{2y}\,\, 
,
\end{array}
\end{equation}
where $d_s \equiv 2\zeta + 2y$, hence 
\begin{equation}\label{expop}
\begin{array}{lcl}
\overline{\left(\frac{\partial K^L_{\mbox{\boldmath\tiny $R$}}}{\partial 
K_i}K_i\right)^2} &\sim& L^{d_s-d}\,\, ,\\[2mm]
\overline{K^L_{\mbox{\boldmath\tiny $R$}}\frac{\partial K^L_{\mbox{\boldmath
\tiny $R$}}}{\partial K_i}K_i} &\sim& L^{2y-d}\,\, ,
\end{array}
\quad\qquad L\gg \xi_-\,\, .
\end{equation}
[We shall write $d_s$ throughout but note that this exponent takes the value 
$d_s=d-1$ within MKRG.] This scaling is due to the fact that bonds 
contribute to the energy fluctuations if and only if they are cut by a 
droplet wall. 
In the limit $T\to 0$, the derivative $\frac{\partial K^L_{\mbox{\boldmath
\tiny $R$}}}{\partial K_i}$ equals $\pm 1$ with probability $\frac{1}{2} 
L^{d_s-d}$ and vanishes otherwise [cf.~(\ref{renK}) in the $\beta\to
\infty$-limit, equation (\ref{approxrg1})]. For the same reason, we obtain 
for the higher moments:
\begin{equation}\label{QLbT}
\begin{array}{lcl}
\overline{\left(\frac{\partial K^L_{\mbox{\boldmath\tiny $R$}}}{\partial 
K_i}K_i\right)^{2q}} &\sim& L^{d_s-d}\,\, ,\\[2mm]
\overline{K^L_{\mbox{\boldmath\tiny $R$}}\left(\frac{\partial K^L_{\mbox{
\boldmath\tiny $R$}}}{\partial K_i}K_i\right)^{2q-1}} &\sim& L^{2y-d}\,\, ,
\end{array}
\quad\qquad L\gg \xi_-\,\, .
\end{equation}
There is therefore {\em no}\/ multifractal but only trivial scaling at 
length scales $L\gg \xi_-$ in the ordered phase. With the aid of (\ref{em}) 
and (\ref{om}) we can write the moments (\ref{QLbT}) in the scaling form
\begin{equation}\label{QLsc}
\begin{array}{lcl}
\overline{\left(\frac{\partial K^L_{\mbox{\boldmath\tiny $R$}}}{\partial 
K_i}K_i\right)^{2q}} &\sim& L^{2\zeta_c^{(q)}-d}\, {\cal F}_q^e(\frac{L}{
\xi_-})\,\, ,\\[3mm]
\overline{K^L_{\mbox{\boldmath\tiny $R$}}\left(\frac{\partial K^L_{\mbox{
\boldmath\tiny $R$}}}{\partial K_i}K_i\right)^{2q+1}} &\sim& L^{y_c+2
\zeta_c^{(q)}-2\zeta_c-d}\, {\cal F}_q^o(\frac{L}{\xi_-})\,\, ,
\end{array}
\end{equation}
with scaling functions that satisfy
\begin{equation}\label{QLscf}
\begin{array}{lcl}
{\cal F}_q^e(0)= 1\,\, ,\\[2mm]
{\cal F}_q^e(x)\stackrel{x\to \infty}{\sim} x^{d_s-2\zeta_c^{(q)}}\,\, ,
\\[2mm]
{\cal F}_q^o(0)= 1\,\, ,\\[2mm]
{\cal F}_q^o(x)\stackrel{x\to \infty}{\sim} x^{2y-y_c-2\zeta_c^{(q)}+2
\zeta_c}\,\, .
\end{array}
\end{equation}
In the limit $d\searrow d_\ell$ along the critical line $T_c(d)$ we know 
that $y_c\searrow 0$ and $\zeta_c\searrow \zeta(d_\ell)$. Since also (
\ref{em}) and (\ref{om}) must then reduce to (\ref{QLbT}), it follows that
\begin{equation}\label{zetampos}
\zeta_c^{(q)} \stackrel{d\searrow d_\ell}{\to} \frac{d_s(d_\ell)}{2} = 
\zeta(d_\ell)\,\, .
\end{equation}
It appears therefore that the zero-temperature exponent $\zeta(d_\ell)$ fans 
out along the critical line into the family of the $\zeta_c^{(q)}$. 

\subsection{Energy fluctuations}\label{ef}

We shall now exploit our extended knowledge of the MKRG to study 
energy-energy fluctuations. These are, generally, obtained as the 
appropriate derivatives of
the logarithm of the partition function 
\begin{equation}\label{Gn}
\log Z = \sum\limits_{\nu=0}^\infty G^{(\nu)}
\end{equation}
with respect to the couplings. Here $\exp G^{(\nu-1)}$ is the multiplicative 
constant that appears in the partition function in the $\nu$th 
renormalisation step. It is the contribution to the free energy $\beta F$ 
coming from excitations on a spatial scale $\ell=2^\nu$. Within the MKRG 
scheme defined by (\ref{renK}) the expression for $G^{(\nu)}$ is
\begin{equation}\label{Gnu}
G^{(\nu)} = \sum\limits_{i_{\nu+1}} g(\{K_{i_\nu}^{(\nu)}\}_{i_{\nu+1}})
\end{equation}
with
\begin{equation}
g(\{K_{i_\nu}^{(\nu)}\}_{i_{\nu+1}}) = \sum\limits_{m=1}^{2^{d-1}}\frac{1}{2}
\log\left[4\cosh(K_{m1}^{(\nu)}+K_{m2}^{(\nu)})\cosh(K_{m1}^{(\nu)}-K_{m2}^{(
\nu)})\right]
\end{equation}
where the sum in (\ref{Gnu}) runs over all blocks $i_{\nu+1}$ of $\nu$th 
level couplings and we have used the convention introduced in section~
\ref{gen} to relabel the $i_\nu$ in an explicit RG step.  

Let
\begin{equation}\label{defencorr}
\Delta e_i\, \equiv\, e_i\, - <e_i>\,\, ,
\end{equation}
where $e_i=s_is_{i'}$, and $s_i$ and $s_{i'}$ are neighbouring spins in the 
original lattice linked by a coupling $K_i$. The simplest energy-energy 
correlation function is then
\begin{equation}\label{defg2}
\beta^2\Gamma_2(r_{ij})\, \equiv\, \beta^2<\Delta e_i \Delta e_j>\, = \,\, 
\sum\limits_{\nu=0}^\infty
K_iK_j\frac{\partial^2 G^{(\nu)}}{\partial K_i\partial K_j}\, .
\end{equation}
A fourth order correlation function of interest is
\begin{equation}\label{fc}
\beta^4\Gamma_4(r_{ij}) \equiv \beta^2[<\Delta e_i^2 \Delta e_j^2> -
<\Delta e_i^2><\Delta e_j^2>] = \sum\limits_{\nu=0}^\infty K_i^2 K_j^2 \frac{
\partial^4
G^{(\nu)}}{\partial K_i^2\partial K_j^2}\, .
\end{equation}

These correlation functions are random quantities. We are now faced with the 
somewhat technical task of calculating averages as well as the averages of 
their powers. The resulting expressions are (\ref{G2r}), (\ref{G4r}), and (
\ref{G2rm}), and their derivation is as follows. 
Let for simplicity $i=i_0$ and $i_1, i_2, \ldots$ denote the locations of 
the couplings on the
successive levels $\nu=1,2,\ldots$ to which $K_i$ contributes and similarly 
for $K_j$. Typically, the couplings $K_{i_\nu}$ and $K_{j_\nu}$ fall in the 
same block
when $\nu$ attains a value $\nu^*(i,j)$ such that
\begin{equation}
2^{\nu^*} = r_{ij} 
\end{equation}
where $r_{ij}$ is the distance between the two spins concerned.
They then contribute to the same $G^{(\nu^*)}$ and to the
same $K_{i_{\nu^*}}^{(\nu^*)} = K_{j_{\nu^*}}^{(\nu^*)}$. Working out (
\ref{defg2}), one finds that the second derivative leads to terms of various 
types. The guiding principle in finding the dominant contribution for large
$r_{ij}$ is that, according to (\ref{em}) and (\ref{om}), averages with
even powers of derivatives are larger than averages with odd powers, while 
some derivatives vanish due to symmetry properties of $g(\{K_{i_\nu}^{(\nu)}
\}_ {i_{\nu+1}})$ and $K_{i_{\nu+1}}^{(\nu+1)}$. Therefore we get
\begin{equation}\label{g2lrij}
\begin{array}{lcl}
\beta^2\Gamma_2(r_{ij}) &\simeq& \frac{\partial^2 G^{(\nu^*)}}{\partial
K_{i_{\nu^*}}\partial K_{j_{\nu^*}}} K_i \frac{\partial
K_{i_{\nu^*}}}{\partial K_i} K_j \frac{\partial K_{j_{\nu^*}}}{\partial K_j} 
+ \\[0.2cm]
{}&{}& +\,\, \sum\limits_{\nu=n^*+1}^\infty \frac{\partial^2 G^{(\nu)}}{
\partial
K^2_{i_\nu}} \prod\limits_{\mu=\nu^*+1}^{\nu-1} \left(\frac{\partial
K_{i_{\mu+1}}}{\partial K_{i_\mu}}\right)^2\frac{\partial^2
K_{i_{\nu^*+1}}}{\partial K_{i_{\nu^*}}\partial K_{j_{\nu^*}}} K_i
\frac{\partial K_{i_{\nu^*}}}{\partial K_i} K_j
\frac{\partial K_{j_{\nu^*}}}{\partial K_j}\\[0.4cm]
\end{array}
\end{equation}
Let $\beta^2\Gamma^L_2(r_{ij})$ be the term with $\nu=n$ in (\ref{defg2}), 
i.e., it is the contribution of scale $L$ fluctuations to $\beta^2
\Gamma_2(r_{ij})$. Obviously, $\Gamma_2^L(r)$ vanishes for $L< r$. Using 
again (\ref{om}) we arrive at 
\begin{equation}
\overline{\Gamma_2^L(r)} \sim
\left(\frac{r}{L}\right)^{d-2\zeta_c}\frac{1}{r^{2d-2y_c}}\,\, , \qquad L
\gaq r\,\, ,
\end{equation}
for the contribution of the energy fluctuations due to two spins a
distance $r$ apart in a volume of linear size $L\gaq r$. 
Hence, summing over length scales, we get the energy-energy correlation 
function at distance $r$:
\begin{equation}\label{G2r}
\overline{\Gamma_2(r)} = \tilde{\sum\limits_{L\gaq r}} \overline{
\Gamma_2^L(r)} \sim
\frac{1}{r^{2d-2y_c}}\,\, ,
\end{equation}
the sum being dominated by the first term.

If one works out the expression (\ref{fc}) for the fourth
cumulant, the fourth derivative leads again to terms of various types. In 
order to find the dominant contribution for large
$r_{ij}$ we proceed as above.
The result is 
\begin{equation}\label{g4lrij}
\begin{array}{lcl}
\beta^4\Gamma_4(r_{ij}) &\simeq& \frac{\partial^4 G^{(\nu^*)}}{\partial^2
K_{i_{\nu^*}}\partial^2 K_{j_{\nu^*}}} K_i^2 \left(\frac{\partial
K_{i_{\nu^*}}}{\partial K_i}\right)^2 K_j^2 \left(\frac{\partial
K_{j_{\nu^*}}}{\partial K_j}\right)^2 \\[0.2cm]
{}&{}&+\,\, \sum\limits_{n=\nu^*+1}^\infty \frac{\partial^4 G^{(\nu)}}{
\partial
K_{i_\nu}^4}\prod\limits_{\mu=\nu^*+1}^{\nu-1} \left(\frac{\partial
K_{i_{\mu+1}}}{\partial K_{i_\mu}}\right)^4\frac{\partial^4
K_{i_{\nu^*+1}}}{\partial K^2_{i_{\nu^*}}\partial K^2_{j_{\nu^*}}} K_i^2
\left(\frac{\partial K_{i_{\nu^*}}}{\partial K_i}\right)^2 K_j^2
\left(\frac{\partial K_{j_{\nu^*}}}{\partial K_j}\right)^2\,\, .
\end{array}
\end{equation}
The quantity $\beta^4\Gamma_4^L(r_{ij})$, defined as the term with $\nu=n$ 
in equation (\ref{fc}), vanishes for $L< r_{ij}$. For $L\gaq r$, averaging 
on the disorder and using (\ref{em}) gives 
\begin{equation}
\overline{\Gamma_4^L(r)} \sim
\left(\frac{r}{L}\right)^{d-2\zeta_c^{(2)}}\frac{1}{r^{2d-4\zeta_c}}\,\, .
\end{equation}
Integrating this equation over length scales $L$ as above, one finds
\begin{equation}\label{G4r}
\overline{\Gamma_4(r)} = \tilde{\sum\limits_{L\gaq r}} 
\overline{\Gamma_4^L(r)} \sim \frac{1}{r^{2d-4\zeta_c}}\,\, ,
\end{equation}
because $d-2\zeta_c^{(2)}$ is the (necessarily positive)\marginpar{disc} 
exponent of a moment of a probability distribution [cf.~also (\ref{em}) and (
\ref{zetampos})].

By taking powers of (\ref{g2lrij}) and averaging with the help of (\ref{em}) 
and (\ref{om}), one finds
\begin{equation}\label{G2rm}
\begin{array}{lcl}
\overline{\Gamma_2^{2q}(r)} &\sim& r^{4\zeta_c^{(q)}-2d}\,\, , \\[2mm]
\overline{\Gamma_2^{2q-1}(r)} &\sim& r^{2y_c+4\zeta_c^{(q)}-4\zeta_c-2d}\,\, 
,
\end{array}
\end{equation}
for the moments of the correlation function $\Gamma_2(r)$. 
Analogous expressions hold for the moments of the higher order correlation 
functions. Since $\zeta_c^{(q)}$, given by (\ref{defzetaq}), is a nonlinear 
function of $q$, this implies {\em multifractal scaling of the
energy-energy correlations at criticality}.

By summing finally (\ref{G2r}) on all $i$ and $j$ in a volume $V$, one 
obtains the mean square fluctuation of the total energy in that volume,
\begin{equation}\label{totef}
\frac{\overline{<\Delta E^2>}}{V} \,\,\sim\,\, \int\limits_{\sim 1}^{V^{
\frac{1}{d}}} dr\,
r^{d-1} \overline{\Gamma_2(r)}\,\,\sim\,\, k_{\mbox{\tiny B}}T_c^2C_V(T_c)\,
\, , \qquad V\to \infty\,\, ,
\end{equation}
valid if $d-2\zeta_c > 0$, which is the case in an interval upward from $d_
\ell$ [cf.~(\ref{em}) and (\ref{zetampos})]. By summing (\ref{G4r}) 
one gets similarly the fourth cumulant of the energy in a volume $V$,
\begin{equation}
\frac{\overline{<\Delta E^4>}_{\mbox{\tiny cum}}}{V}\,\, \equiv\,\, \frac{
\overline{<\Delta E^4>} - 3 \overline{<\Delta E^2>^2}}{V}
\end{equation}
The result depends on the sign of $4\zeta_c - d$ and reads
\begin{equation}\label{4thcres}
\begin{array}{lcl}
\frac{\overline{<\Delta E^4>}_{\mbox{\tiny cum}}}{V} &\sim&
\int\limits_{1}^{V^{\frac{1}{d}}}\, dr\, r^{d-1}\overline{\Gamma_4(r)} 
\\[0.5cm]
{}&\stackrel{V\to \infty}{\sim}& \left\{ \begin{array}{r@{\qquad\mbox{if}
\quad}lr}
V^{4\frac{\zeta_c}{d}-1} & 4\zeta_c/d > 1\,\, . \qquad&\mbox{(a)} \\
  \mbox{cst.} & 4\zeta_c/d < 1\,\, . \qquad&\mbox{(b)} 
\end{array} \right.
\end{array}
\end{equation}
Equation (\ref{4thcres}a) is likely to hold in a narrow range of dimensions
upward from $d_\ell$, and equation (\ref{4thcres}b) for higher dimensions.
Equations (\ref{totef}) and (\ref{4thcres}b) have non\-ana\-lytic correction 
terms in $V$. In nonrandom thermodynamic systems the energy cumulants tend 
to constants in the thermodynamic limit. Equation (\ref{4thcres}a), in 
contrast, expresses an anomalous scaling of the fourth cumulant with volume. 
It is easy 
to check that the same is true for the higher order cumulants.
Such an anomalous scaling of the energy cumulants in spin 
glass was first pointed out by FH for the ordered phase [see \cite{FH}, 
equations (7.11) and (7.12)]. The basic reason for this anomalous behaviour 
of the cumulants is that the energy fluctuations are collective, even 
beyond the scale $\xi_-$, up to scale infinity, and involve larger and 
larger energy differences.

\setcounter{equation}{0}
\section{Geometric interpretation of the critical scaling laws}\label{pi} 

Our purpose is now to understand the scaling behaviour found in the previous
section in terms
of a geometric picture. In the ordered phase, for length scales $L\gg \xi_-$,
the work of Fisher and Huse \cite{FH,FHua} has made it clear how to pass
from MKRG to such a picture. After recalling how this is done, in 
section~{Md}, we will address the same issue in the critical region, 
in section~\ref{Mc}.

\subsection{MKRG and the droplet picture of the ordered phase}\label{Md}

\subsubsection{Droplets}

At length scales $L\gg \xi_-$ the block spins of the MKRG procedure are
naturally identified with the droplets of the FH picture.
Thermally active block spins, i.e., block spins coupled to their 
environment with energies less than $k_{\mbox{\tiny B}}T$, then represent 
thermally 
active droplets. The reversal energy of a scale-$L$ block spin
is equal to twice the sum of the couplings $K^L_{\mbox{\boldmath\tiny $R$}}$ 
that link it to its environment (with signs 
equal to the values of the neighbouring block spins). 
We shall speak of the $K^L_{\mbox{\boldmath\tiny $R$}}$ as the excitation 
free energies of
collective spin reversals on scale $L$, even though strictly speaking
it is always the sum of a few $K^L_{\mbox{\boldmath\tiny $R$}}$ 
that is involved.

The probability
distribution of the $K^L_{\mbox{\boldmath\tiny $R$}}$ scales with $L$ in the 
same way as the probability distribution $P_L$ of 
the reversal free energies $\Delta F^L_{\mbox{\boldmath\tiny $R$}}$ of the 
FH droplets, viz.~as
\begin{equation}\label{P}
P^L(\Delta F^L_{\mbox{\boldmath\tiny $R$}}) \simeq {1\over \Upsilon (T)L^y} {
\cal P}\left({\Delta F^L_{\mbox{\boldmath\tiny $R$}} \over \Upsilon (T)L^y}
\right)\,\, , \quad\qquad L \to \infty\,\, .
\end{equation}
Here $\Upsilon(T)$ is the temperature dependent stiffness coefficient of the 
ordered phase with $\Upsilon(0)\sim J$ and $\Upsilon(T)\stackrel{T\nearrow 
T_c}\sim \xi_-^{-y}$ \cite{CBC,FH}.
Equation (\ref{P}) implies in particular that the probability $f^L(T)$ for 
a scale $L$ block spin to be thermally active is 
\begin{equation}
f^L(T) \sim  \frac{k_{\mbox{\tiny B}}T}{\Upsilon(T)L^y}\,\, ,\qquad L\gg\xi_-
\,\, .  
\end{equation}
Therefore in the large-$L$ limit the thermally active scale $L$ block spins
become infinitely dilute and may be considered as independent degrees
of freedom.

FH argue that the droplet wall has a fractal dimension $d_s$ with 
$d-1<d_s<d$. 
At low enough temperature the droplet wall will be of zero width
and cut through a well-defined set of couplings of the unrenormalised system.
The associated reversal
energy is the sum of the energies of each of the couplings cut.
The number of couplings cut will grow typically as $L^{d_s}$.
MKRG gives $d_s=d-1$. This is easily seen if one
expands the elementary renormalisation step (\ref{renK}) in the 
$\beta\to\infty$ limit. It reads
\begin{equation}\label{approxrg1}
K_{i_{n+1}}^{(n+1)}\,\, =\,\, \sum\limits_{m=1}^{2^{d-1}} \mbox{sgn}
\left(K^{(n)}_{m1}K^{(n)}_{m2}\right)
\min\left(|K^{(n)}_{m1}|,|K^{(n)}_{m2}|\right)
\end{equation}
at $T=0$. Upon differentiating (\ref{approxrg1}) with respect to the $2^d$
couplings that enter into the expression one finds that, at {\em zero 
temperature}\/, the derivatives
with respect to $2^{d-1}$ of them are equal to $\pm 1$, whereas the others
vanish. After $n$ renormalisation steps with $2^n=L$ one sees
that the scale-$L$ block spin corresponds to a droplet with wall area $\sim 
L^{d-1}$. This special value for $d_s$, i.e., $d_s=d-1$, must nevertheless 
be ascribed
to the particular choice of the MKRG recursion relations and is not
fundamental. Numerical work \cite{BMds2125,CMBds212,CMBds231225,Hds322} 
points to nontrivial values of $d_s$ and one can envisage an improved MKRG 
that would lead to 
$d_s\ne d-1$. The wall of a droplet at {\em finite temperature}\/ has an 
intrinsic width of the order of the correlation length $\xi_-$ (see 
figure~3).
But since the RG flow carries the system to the zero temperature
fixed point, the droplet walls coarse grained to at least the scale $\xi_-$ 
have the same fractal dimension $d_s$ at length scales $L\gg \xi_-$.
\begin{figure}[htb]
\vspace{0.5cm}
\parbox[b]{15.5cm}{${}$\hspace{6.9cm} ${}$}
\vspace{4cm}
\centering{
\begin{picture}(13,5)
\put(7,4.5){\makebox(0,0){\small This figure may be requested from the
authors}}
\end{picture}
}
\vspace{2cm}

\caption{A scale-$L$ droplet at finite temperature ($L\gg\xi_-$).
The wall of a droplet at {\em finite temperature}\/ has an 
intrinsic width of the order of the correlation length $\xi_-$.
In the large-$L$ limit, scale-$L$ thermally active droplets get more and 
more 
dilute and therefore may be treated as noninteracting.
By the same token, it consists of one and only
one main connected domain of order $L^d$ spins, whose wall width is 
negligible with respect to its own linear size. Therefore droplets 
are two-level systems, with ``up" and ``down" referring to the main domain.
}
\end{figure}

In the large-$L$ limit, scale-$L$ thermally active droplets get more and 
more 
dilute and therefore may be treated as noninteracting.
By the same token, upon assuming that there are no correlations between
the spatial positions of thermally active droplets of different scales,
one may neglect interactions between scale-$L$ droplets and
droplets on scales just below (e.g., $L/2$ and $L/4$).
As one goes down further in scale, however, droplet-droplet interactions
will at some point come into play, as may be estimated as follows.
On some arbitrary linear scale $\ell<L$, the wall of a scale-$L$ droplet
will pass through $(L/\ell)^{d_s}$ blocks of size $\ell$ of which 
typically a number $f^\ell(T) (L/\ell)^{d_s}$ will contain thermally active 
droplets interacting with the wall. For $L$ below a value that following FH 
\cite{FH} we shall 
call $\lambda_L$, this number will typically be larger than unity and the 
wall of the scale-$L$ droplet is ``interaction widened" up to the scale $
\lambda_L$. An expression for
$\lambda_L$ can readily be found \cite{FH} but of interest here is only the 
property 
\begin{equation} 
\lim\limits_{L\to\infty}\frac{\lambda_L}{L} = 0\,\, .
\end{equation}
It implies that for large $L$ a droplet consists of one and only
one main connected domain of order $L^d$ spins, whose wall width is 
negligible with respect to its own linear size. Therefore droplets 
are two-level systems, with ``up" and ``down" referring to the main domain.

\subsubsection{Interaction between droplets}

How does the interaction between droplets of different
scales appear in the MKRG formalism ?
To answer that question we consider a string of two scale-$L$ couplings 
$K^{(n)}_1$ and $K^{(n)}_2$ (contributing to a higher level coupling 
$K_{i_{n+1}}^{(n+1)}$) with in the center
a scale-$L$ block spin $S_0$ and at the ends two larger scale block 
spins $S_1$ and $S_2$ (one of them is necessarily a
scale-$2L$ spin and the other one belongs to a scale $\gaq 4L$).
The derivative of (\ref{approxrg1}) with respect to a scale-$L$ 
coupling $K_{mk}^{(n)}$ is bounded by
\begin{equation}\label{der}
0 < \left|\frac{\partial K_{i_{n+1}}^{(n+1)}}{\partial K^{(n)}_{mk}}\right| 
< 1\,\, .
\end{equation}
The preceding discussion implies that when its absolute value is not 
close to the lower or upper limit values $0$ or $1$, 
respectively, this is the signature of thermal activity. 
If this is true for the derivative with respect to $K_1^{(n)}$ or 
$K_2^{(n)}$, 
then the wall between the droplets $S_1$ and $S_2$ may cut 
either bond of the string. The two positions of the wall differ 
by the orientation of $S_0$. Hence the droplet $S_0$ thermally 
decorates the wall between the droplets $S_1$ and $S_2$.
The two wall positions will in general have different weights.
Upon working out the derivatives of interest with the aid of 
(\ref{renK}) we find  
\begin{equation}\label{dK}
\frac{\partial K_{i_{n+1}}^{(n+1)}}{\partial K^{(n)}_{i}} = \frac{\frac{1}{2}
\sinh (2K^{(n)}_{i})}{\cosh\left(K^{(n)}_{1}+K^{(n)}_{2}\right)\cosh
\left(K^{(n)}_{1}-K^{(n)}_{2}\right)}\,\, , \qquad i=1,2\,\, . 
\end{equation}
The two derivatives are of comparable absolute value, and 
hence thermal activity is nonnegligible, only when 
\begin{equation}\label{2s}
\left|K^{(n)}_{1} \pm K^{(n)}_{2}\right| \la \frac{k_{\mbox{\tiny B}}T}{J}\,
\, .  
\end{equation}
The $\pm$ sign corresponds to the two possible relative orientations
of $S_1$ and $S_2$. When (\ref{2s}) is satisfied for one of the two signs, 
it 
will not be, in general, for the other sign. This is the ``interaction"
between droplets of different scales. This feature of the MKRG
also appears in the FH droplet picture: a scale $\ell$ droplet lies on the
wall of a scale $L$ droplet ($L>\ell$), has a wall section of linear size $
\ell$ in common
with it, and is thermally active for one, but not for the other orientation
of the large droplet.

At low critical temperatures the 
derivative (\ref{dK}) has a probability distribution on the interval
$[-1,1]$ which {\em strongly peaks}\/ at $0$ and $\pm 1$ and is small 
elsewhere. 
At $T=0$, it is the sum of three {\em delta peaks}\/ at $0$ and $\pm 1$.
Since the ordered phase is controlled by the zero temperature
fixed point, the phenomenon of interaction widening discussed here
plays no r\^ole for the large scale properties in the spin glass ordered
phase. It is important, however, for the understanding of the crossover from 
zero temperature to criticality, as we shall see below.
  
\subsection{The low-dimensional critical region - protodroplets}\label{Mc} 

\subsubsection{Surfaces}

In the critical region, that is, on length scales $L$ below the 
correlation length and whether below or above $T_c$, the FH droplets 
cannot be defined. Yet,
in the MKRG picture thermally active block spins can be defined as before, 
viz.~by the condition that
their interaction energy with the environment be less than some
constant times $k_{\mbox{\tiny B}}T$. 

Let us consider for definiteness the critical line in the $T$-$d$ plane.
On this line the probability for a block spin to be thermally active
is of the order of $k_{\mbox{\tiny B}}T_c/J \equiv p > 0$, whatever the 
spatial scale $L$. Therefore there must be interactions 
between these spins whose importance is characterised by the parameter $p$.
Similarly, from the fact that along the critical line the derivatives 
(\ref{dK}) have a nontrivial probability distribution it 
follows that there are interactions between thermally active block spins of 
neighbouring scales ($L$ and $2L$). 

Block spins can no longer be identified with droplets and we should look
for a new interpretation. The key to this interpretation is to consider
the limit $p\to 0$.
For $d$ very close to $d_\ell$ the scale independent probability $p$ for a 
block spin to be thermally active is very small.
Also in this limit the probability law for the derivatives (\ref{dK})
approaches the trivial three-peak distribution,
which means that the interaction between block spins of neighbouring
scales disappears.

Let us now look at the unrenormalised lattice (scale $L=1$) 
for $p$ extremely small. The thermally active spins will be extremely
dilute so that for the moment we may ignore the interactions in the 
discussion. 
We could indicate the thermally active spins 
in a figure by surrounding them with a {\em surface}\/.
For $L$ not too large we now identify a thermally active scale-$L$ block
spin with a size-$L$ surface that encloses a {\em cluster}\/ of 
unrenormalised
spins that may reverse collectively with respect to their environment
due to thermal agitation. 
These larger surfaces may be irregularly shaped.  
The essential point is that the surfaces of the individual thermally 
active spins as well as of the thermally active clusters 
are of width zero, because not interacting with any other thermally
active degree of freedom. These thermally active surfaces are systems 
with two energy levels.
They may be considered as independent degrees of freedom under certain 
circumstances, e.g., in the limit $p\ll 1$ and/or for not too large 
values of $L$, or for the calculation of the spin-spin correlation 
function (see section~\ref{ssc}).
 
\subsubsection{Interaction between surfaces}

We now ask about the effect of interactions between the surfaces. 
Interactions occur due to surface sections 
that they have in common. In the small-$p$ limit, the thermally active
surfaces occupy random positions in space. Let us consider two
scales $L$ and $\ell<L$. If $\overline{d}(T_c)$ is the surface fractal
dimension, then a size-$L$ surface will pass through $(L/\ell)^{
\overline{d}(T_c)}$ regions
of linear size $\ell$ and interact typically with a number $p(L/\ell)^{
\overline{d}(T_c)}$
of size $\ell$ thermally active surfaces. This number becomes $\ga 1$ for
\begin{equation}
L \ga L_{\mbox{\tiny entr}} \equiv \,\, \left(\frac{1}{p}\right)^\frac{1}{
\overline{d}(T_c)}\,\, \sim\,\, \left(\frac{J}{k_{\mbox{\tiny B}}T_c}\right)^
\frac{1}{\overline{d}(T_c)}\,\, .
\end{equation}
Hence beyond this length scale the surface entropy plays a r\^ole, i.e., the 
thermally active surfaces are likely to
interact with smaller size thermally active surfaces with which they
then form what we shall call {\em composite thermally active surfaces}\/.

The probability for a thermally active surface to interact with another
one of size comparable to its own is small (of order $p$), but
is also taken into account at this level of approximation. 
We should therefore ask if the interactions do not lead to a coupling,
direct or indirect, between all thermally active degrees of freedom in
the lattice. It is not difficult to estimate that for small enough $p$
the interactions couple the thermally active degrees of freedom only
into finite disjoint sets.
We call such a set an {\em isolated thermally active composite surface}\/ 
(see figure~4).
An appropriate name which is much shorter is a {\em protodroplet}\/, as will
appear in subsection~\ref{summ}.
\begin{figure}[htb]
\setlength{\unitlength}{1cm}
\begin{picture}(13,8.5)
\put(7,4.5){\makebox(0,0){\small This figure may be requested from the
authors}}
\end{picture}

\caption{A scale-$L$ protodroplet in the critical region. The composite 
thermally active surfaces are {\em many}\/-level systems
of coupled degrees of freedom. Since the composing surfaces may be of
comparable size, the protodroplet cannot be simply characterised
by a single up/down variable. It is a hierarchical object, composed of
interacting two-level systems - the surfaces - which is statistically 
self-similar up to the scale of its own size. The energy fluctuations are 
due to thermally active bonds between adjacent spins that are 
cut by the surfaces, as indicated in the figure. The multifractality 
of the energy correlations is a
consequence of the nontrivial distribution of relative weights between
the two possible orientations of each composing surface.}
\end{figure}

The composite thermally active surfaces are {\em many}\/-level systems
of coupled degrees of freedom. Since the composing surfaces may be of
comparable size, the protodroplet cannot be simply characterised
by a single up/down variable. It is a hierarchical object, composed of
interacting two-level systems - the surfaces - which is statistically 
self-similar up to the scale of its own size. Its multifractality is a
consequence of the nontrivial distribution of relative weights between
the two possible orientations of each composing surface. If the above 
argumentation, starting from thermally active surfaces and
leading to the identification of protodroplets, is supplemented with
appropriate scaling hypotheses, it may be seen as an alternative  -
valid in the limit $d \searrow d_\ell$ - to the MKRG formalism in 
providing a basis for the theory of the low-dimensional critical state. 

\subsection{Summary}\label{summ}

On the critical line $T_c(d)$ in the temperature-dimension plane the spin 
glass system is statistically invariant under scale transformations, and has 
thermal 
fluctuations on all length scales. The equilibrium ensemble is not 
dominated by a single representative configuration or class of 
configurations, 
which at first sight renders a geometric analysis forbiddingly difficult.
Such an analysis is nevertheless possible, for sufficiently low $T_c(d)$, 
i.e., for $d$ close to $d_\ell$. The simplifying feature is that at low 
temperatures, and whether the system is critical or not, the fraction of 
couplings that effectively participate in the energy fluctuations goes to 
zero. The thermally active couplings therefore determine a set of surfaces 
of low spatial density. Each surface encloses a cluster of spins which may 
reverse with respect to a frozen local environment {\em in the ordered phase 
as well as at criticality}\/. 

What has been said in section~\ref{Mc} holds at $T=T_c$ 
as well as in the whole critical region. Imagine now that for some $T$ 
below $T_c$ we increase the length scale $L$. The probability for the
protodroplet to be composed of two comparable size surfaces will
go down from its value $p$ for $L\la\xi_-$ to zero for $L\to\infty$.
For $L\ga\xi_-$ a single surface will dominate and the protodroplet 
has become a droplet. Hence the concept of protodroplets generalises 
the concept of droplets in dimensions near $d_\ell$ and permits a 
geometrical picture that also applies to the critical region. Random
reversals of protodroplets are at the origin of chaos in the ordered phase
as well as at criticality.

\setcounter{equation}{0}
\section{Characteristic lengths near the lower critical dimension}
\label{char}

Near the lower critical dimension there is a confluency of zero-temperature 
and critical phenomena. In order to clarify the picture, we analyse in this 
subsection the characteristic lengths that play a r\^ole, in particular the 
correlation length $\xi_-$ and the overlap length $L_{\Delta T}$ defined in (
\ref{ollT}).

\subsection{Correlation length near $d_\ell$}

The correlation length above and below $T_c$ is determined by the RG 
transformation ${\cal R}(K)$, whose explicit form near $d_\ell$, equation (
\ref{renKlT}), can also be written as
\begin{equation}
K^{2L}\,\, =\,\, 2^y\,K^L - c_0\, (K^L)^{-1} + \cdots\,\, ,
\end{equation}
where $c_0$ is a positive constant. For further analysis it is convenient to 
pass to continuous $L$, which gives
\begin{equation}\label{KlogL}
\frac{dK^L}{d\log L}\,\, =\,\, y\,K^L - c\,(K^L)^{-1} + \cdots
\end{equation}
with $c$ a positive constant. At low temperatures we may neglect the dot 
terms in (\ref{KlogL}). Above $d_\ell$, the exponent $y(d)$ is positive and 
equation (\ref{KlogL}) has the two fixed-point solutions $K^L=\infty$ and 
$K^L=\sqrt{c/y} \equiv K_c$. The latter relation can be recast as
\begin{equation}\label{ylT}
y = c \left(\frac{k_{\mbox{\tiny B}}T_c}{J}\right)^2\,\, .
\end{equation}
Since $y$ is linear in $d-d_\ell$, this shows that $T_c\sim \sqrt{d-d_\ell}$ 
just above $d_\ell$ \cite{SY,FH}. At low temperatures, the solution of (
\ref{KlogL}) for initial value $K^{(0)}=K=\frac{J}{k_{\mbox{\tiny B}}T}$ is
\begin{equation}\label{lcRG}
\frac{(K^L)^2}{K_c^2} - 1\,\, =\,\, \left(\frac{K^2}{K_c^2} - 1\right)\, 
L^{2y}\,\, .
\end{equation}
This shows that just above $d_\ell$ the critical point exponent $y_c$ is 
equal to $2y$, as also found by \cite{MM,FH}. Correlation lengths $\xi_+(T)$ 
above $T_c$ and $\xi_-(T)$ below $T_c$ may be defined by
\begin{equation}\label{defxi}
\begin{array}{lcl}
(K^{\xi_+})^{-1} &=& {\mit f_+}\,\, ,\\[2mm]
(K^{\xi_-})^{-1} &=& {\mit f_-}\,K_c^{-1}\,\, ,
\end{array}
\end{equation}
where ${\mit f_+}$ is a fixed constant often set equal to unity and ${\mit 
f_-}$ a fixed constant less than $1$. These are the usual definitions; in 
particular, at $(K^{\xi_-})^{-1}={\mit f_-}\,K_c^{-1}$ the renormalisation 
group crosses over from critical to zero temperature fixed point control. 
From (\ref{lcRG}) and (\ref{defxi}) we find 
\begin{equation}
\begin{array}{lclc}
\xi_+&=&\left[1-\left(\frac{k_{\mbox{\tiny B}}T_c}{{\mit f_+}\,J}\right)^2
\right]^\frac{1}{2y}\left[1-\left(\frac{T_c}{T}\right)^2\right]^{-
\frac{1}{2y}}\,\, ,\qquad &T>T_c\,\, ,\\[2mm]
\xi_-&=&\left[\frac{1}{{\mit f_-}^2}-1\right]^\frac{1}{2y}\left[\left(
\frac{T_c}{T}\right)^2-1\right]^{-\frac{1}{2y}}\,\, ,\qquad &T<T_c\,\, .
\end{array}
\end{equation}

Some comments are in place. Using (\ref{ylT}), we find that for $d\searrow d_
\ell$ the coefficient of $\xi_+$ has the well-defined limit
\begin{equation}
\left[1-\left(\frac{k_{\mbox{\tiny B}}T_c}{{\mit f_+}\,J}\right)^2\right]^
\frac{1}{2y} \sim e^{\frac{J^2}{2c(k_{\mbox{\tiny B}}T_c)^2}\log\left(1-
\frac{(k_{\mbox{\tiny B}}T_c)^2}{({\mit f_+}\,J)^2}\right)} \simeq e^{-
\frac{1}{2c{\mit f_+}^2}}\,\, ,\qquad d\searrow d_\ell\,\, .
\end{equation}
Hence the definition (\ref{defxi}) fixes $\xi_+$ near $d_\ell$ up to a 
finite multiplicative constant. At $d_\ell$, we have
\begin{equation}\label{dlcl}
\xi_+(T) \simeq e^{-\frac{1}{2c{\mit f_+}^2}+\frac{J^2}{2c(k_{\mbox{\tiny 
B}}T)^2}}\,\, ,
\end{equation}
as also found by McMillan \cite{MM} and Bray and Moore \cite{BMmf}. If, 
however, we try to take the limit $d\searrow d_\ell$ with $\frac{T}{T_c}$ 
fixed in the expression for $\xi_-$, we find
\begin{equation}\label{xi-p}
\lim\limits_{{d\searrow d_\ell}\atop{\frac{T}{T_c}\, \mbox{\tiny fixed}}} 
\xi_-(T) = \lim\limits_{y\searrow 0}\left(\frac{\frac{1}{{\mit f_-}\,^2}-1}{
\frac{T_c^2}{T^2}-1}\right)^\frac{1}{2y} = \left\{\begin{array}{llcl}\infty
\, ,\qquad& {\mit f_-}&<& \frac{T}{T_c}\\ 0\, ,\qquad & \frac{T}{T_c}&<&{
\mit f_-}\end{array}\right.\,\, ,
\end{equation}
so that the limit value of $\xi_-$ at $\frac{T}{T_c}$ fixed is $0$ or 
$\infty$ depending on the arbitrary constant ${\mit f_-}\,$. Of course, 
in the $T$-$d$ plane, the point ($T=0$, $d=d_\ell$) lies on the 
intersection of two lines, with $\xi_-=0$ and $\xi_-=\infty$ so 
that $\xi_-(T,d)$ has a mathematical singularity in ($0,d_\ell$). 
Nevertheless this feature is insignificant if one keeps $d$ fixed. At
fixed $d$, equation (\ref{defxi}) remains sensible. Equation (\ref{xi-p})
only indicates that the range of length scales $L$ at which one is uncertain
whether being below or above $\xi_-$ becomes larger upon lowering the 
dimension towards $d_\ell$.

\subsection{Chaos near the lower critical dimension}

Chaos in the critical region has first been shown to exist by a RG
analysis in which the density of zeros of a scale-$L$ coupling on
the temperature axis was considered. In this subsection we will
illucidate in more detail the critical chaos near the lower critical
dimension and show how it is reflected in the change in activity of
the thermal degrees of freedom.

\subsubsection{The density of zeros of a scale-$L$ coupling on the $T$-axis}

It is of interest to consider the autocorrelation function $C^L(T,T+\Delta 
T)$ on the temperature axis of a renormalised coupling at scale $L$. For 
this quantity, normalised to $1$ at $\Delta T=0$, one expects the behaviour
\begin{equation}
1-C^L(T,T+\Delta T) \simeq \frac{1}{2}\Delta T^2 \rho_L^2(T)\,\, ,\qquad 
\mbox{as $\Delta T\to 0$}\,\, ,
\end{equation}
where $\rho_L(T)$ is the {\em average density of zeros}\/ of such a 
coupling. This quantity is, in principle, determined when a renormalisation 
group transformation is given. 

Below we will calculate $\rho_L(T)$ near the lower critical dimension. 
Within the approximation of the Migdal-Kadanoff renormalisation group, 
Ney-Nifle and Hilhorst \cite{NH1,NH2} found an expression for $\rho_L(T)$ 
which slightly rewritten reads
\begin{equation}\label{recCn}
T^2 \rho_L^2(T)\,\, =\,\, K^2\,\sum\limits_{j=0}^{n-1}\, \left(
\frac{dK^{(j)}}{dK}\right)^2\, \mu(K^{(j)})\lambda(K^{(j+1)})\ldots
\lambda(K^{(n-1)})\,\, .
\end{equation}
Here, as before, $K=\frac{J}{k_{\mbox{\tiny B}}T}$; the functions $\lambda$ 
and $\mu$ are determined by the renormalisation group with $\lambda(K) 
\equiv 2^{z(K)}$ increasing from $2^{\zeta_c}$ to $2^{\zeta}$ in the 
interval ($K_c$, $\infty$) and $\mu(K)$ behaving as $\mu(K)\simeq\frac{\mu_
\infty}{K^5}$ for $K\to\infty$, where $\mu_\infty$ is a constant whose MKRG 
value is $\approx 0.12$\,.

Passing to a continuous length scale and using (\ref{lcRG}) for the 
$K^{(i)}$ gives
\begin{equation}
T^2\rho_L^2(T) = \frac{\mu_\infty}{\log 2} \frac{ K^4}{K_c^7} \int\limits_1^L
\, \frac{d\ell}{\ell}\, \left[1+\left(\frac{K^2}{K_c^2}-1\right)\ell^{2y}
\right]^\frac{7}{2}\,\, \exp\left(2\int\limits_\ell^L\, \frac{d{\tilde{
\ell}}}{{\tilde{\ell}}}\, z(K^{\tilde{\ell}})\right)\,\, .
\end{equation}
In the limit of interest we can expand $z(K)$ around $K=\infty$ in powers of 
$K^{-1}$. From equation (5.5a) of \cite{NH2} we can deduce that $z$ is even 
in $K$ so that to lowest nontrivial order
\begin{equation}\label{devzeta}
z(K) = \zeta - \frac{1}{2}\zeta''K^{-2}\,\, .
\end{equation}
Using (\ref{devzeta}) and again (\ref{lcRG}), and setting $\tau^{-1} \equiv 
\frac{T_c^2}{T^2}-1$, we obtain 
\begin{equation}
\rho_L^2(T) \stackrel{d\searrow d_\ell}{\simeq} \frac{\mu_\infty}{\log 2}
\left(\frac{k_{\mbox{\tiny B}}}{J}\right)^3 \frac{ T_c^7}{T^6} \int
\limits_1^L\, \frac{d\ell}{\ell}\, \left[1+\tau^{-1}\ell^{2y}\right]^
\frac{7}{2}\, \exp\left(2\int\limits_\ell^L\, \frac{d{\tilde{\ell}}}{{\tilde{
\ell}}}\, \left[\zeta - \frac{1}{2}\zeta''K_c^{-2} \left[1+\tau^{-1}{\tilde{
\ell}}^{2y}\right]^{-1}\right]\right)\,\, .
\end{equation}
Changing variables to $s\equiv \frac{\ell^{2y}}{\tau}$, we get finally for 
the average density of zeros on the temperature axis
\begin{equation}\label{rhores}
\rho_L^2(T) \stackrel{d\searrow d_\ell}{\simeq} \frac{\mu_\infty}{\log 2}
\left(\frac{k_{\mbox{\tiny B}}}{J}\right)^3 \frac{T_c^7}{T^6} \left(
\frac{L^{2y}}{\tau}\right)^\frac{\zeta}{y} \left(1+\frac{\tau}{L^{2y}}
\right)^\frac{\zeta''}{2c} I_y\left(\frac{L^{2y}}{\tau}\, ,\, \frac{1}{\tau}
\right)
\end{equation}
with
\begin{equation}\label{Iy}
I_y \equiv \frac{1}{2y}\,\int\limits_\frac{1}{\tau}^\frac{L^{2y}}{\tau}\, ds
\, s^{-1 - \frac{\zeta}{y} + \frac{\zeta''}{2c}}\, (1+s)^{- \frac{7}{2} - 
\frac{\zeta''}{2c}}\,\, .
\end{equation}
We are now in a position to consider various limits. From (\ref{rhores}) and 
(\ref{Iy}) we get
\begin{equation}\label{rhoresl1}
\rho_L(T) \stackrel{d\searrow d_\ell}{\simeq} \left\{ 
\begin{array}{lrr}
\sqrt{\frac{\mu_\infty}{2\zeta\log 2}}\sqrt{\frac{k_{\mbox{\tiny 
B}}^3T}{J^3}} L^{\zeta}\,\, ,\quad &\mbox{ $T\to 0$}\,\, . \qquad&\mbox{(a)} 
\\
\sqrt{\frac{\mu_\infty}{2\zeta_c\log 2}}\sqrt{\frac{k_{\mbox{\tiny 
B}}^3T_c}{J^3}} L^{\zeta_c} \left[1 + \left|\frac{T-T_c}{T}\right|\left(
\frac{L^{2y}\zeta''}{2c}-\frac{1}{2}\right)+ {\cal O}(\left|\frac{T-T_c}{T}
\right|^2) \right]\,\, ,\quad &\mbox{ $\left|\frac{T}{T_c}-1\right|\ll 1$}\,
\, . \qquad&\mbox{(b)} 
\end{array} \right.
\end{equation}
The expression (\ref{rhoresl1}b) shows that for large enough $L$ the density 
of zeros has a negative derivative at $T_c$, and must therefore have a 
maximum between $T=0$ and $T=T_c$. For smaller $L$, the derivative at $T_c$ 
is positive and the maximum of $\rho_L(T)$ must occur for $T>T_c$. The 
crossover value is
\begin{equation} L\stackrel{d\searrow d_\ell}{\simeq} \exp\left(\frac{\log(1+
\frac{c}{\zeta''})}{2y}\right)
\end{equation}
In the limit $d\searrow d_\ell$ at $L$ and $T$ fixed, we find
\begin{equation}\label{rhoresl2}
\rho_L(T) \stackrel{d\searrow d_\ell}{\simeq} \sqrt{\frac{\mu_\infty}{2\zeta
\log 2}}
\sqrt{\frac{k_{\mbox{\tiny B}}^3T}{J^3}} L^{\zeta}\,\, ,
\end{equation}
valid in the whole interval of temperatures $0\laq T\laq T_c$. 
The $\sqrt{T}$ behaviour at low temperatures was known for 
$d>d_\ell$ \cite{FH,NH1,NH2} and for $d<d_\ell$ \cite{NH1,NH2}. 
The present calculation shows that this behaviour persists when 
$d$ passes through $d_\ell$.

\subsubsection{Confluency of critical and zero-temperature phenomena
near $d_\ell$}\label{confl}

A last comment on the confluency of critical and zero temperature properties 
within the geometric picture near $d_\ell$ is in place. Each sign reversal 
of a coupling $K^L_{\mbox{\boldmath\tiny $R$}}$ on the temperature axis will 
typically be accompanied by a sign reversal of the correlation $<s_0s_L>$ 
between two spins a distance $L$ apart in the region around $\mbox{\boldmath 
$R$}$. From (\ref{rhoresl2}) for the total number of sign reversals of 
$<s_0s_L>$ between $0$ and $T_c$
\begin{equation}
\begin{array}{lcl}
{\cal R}(L) &=& \int\limits_0^{T_c}dT \rho_L(T)\\
{}&\sim& J^{-\frac{3}{2}}T_c^\frac{3}{2}L^\zeta\,\, ,
\end{array}
\end{equation}
valid near $d_\ell$. So, typically, as one passes from the ground state to 
the critical state there are no sign reversals of $s_0$ with respect to 
$s_L$ below a length scale $L_{\mbox{\tiny gs}}$ given by
\begin{equation}
L_{\mbox{\tiny gs}} \sim \left(\frac{J}{k_{\mbox{\tiny B}}T_c}\right)^
\frac{3}{2\zeta}\,\, .
\end{equation}
Hence below this length scale, a protodroplet around a given point $\mbox{
\boldmath$R$}$ at criticality coincides {\em typically}\/ with the droplet 
at zero temperature in that region, even though an average over the whole 
system yields critical scaling and statistical invariance with length scale.

\setcounter{equation}{0}
\section{Spin-spin correlations and response to a magnetic field}\label{ssmf}

\subsection{Spin-spin correlation functions}\label{ssc}

We wish to consider here spin-spin correlations at
criticality. Instead of using the MKRG formalism as we did for the
energy-energy correlations, we shall now make freely use of the
interpretation in terms of thermally active surfaces developed in
section~\ref{Mc}. For the following calculation we consider the
thermally active surfaces as independent two-level systems, i.e.,
we neglect their interactions.

Let  $<s_is_j>_c  = <s_is_j> - <s_i><s_j>$  be the connected correlation
function between two spins a distance  
$L = 2^n$ apart. In this section $i$ and $j$ denote spin sites and not,
as before, bonds. By a ``surface" enclosing a given spin at a given scale we
mean in what follows the one with the lowest excitation {\em energy}\/ 
of all possible such surfaces; it need not, and in general will not, 
be thermally active. With respect to the thermally active surfaces 
introduced above these surfaces are exactly what the droplets are to 
the thermally active droplets.
 
Let $k$ be the index denoting the surface of scale $2^k$ that encloses spin 
$i$
but not spin $j$, and let similarly $l$ be the index of the scale $2^l$
surface that encloses spin $j$ but not spin $i$. Furthermore let $m$ be the
index of the scale $2^m$ surfaces that enclose both spins.
We shall denote the excitation energies of these surfaces by $\epsilon_k$,
$\epsilon_l$, and $\epsilon_m,$ respectively.

The spin clusters enclosed by the surfaces introduced here can now be
treated in the same way as the FH droplets in the ordered phase 
\cite{FH,FHua}. For the spin-spin correlation it is sufficient to work in the
approximation of independent surfaces; taking their interaction into
account would only renormalise several of the constants but not lead to
any new effects.

Upon treating the surfaces as independent two-level systems one finds
\begin{equation}
<s_is_j>_c \simeq s_i^0s_j^0\, \prod\limits_{k=1}^{n-1} \tanh\frac{\beta
\epsilon_k}{2}\, \prod\limits_{l=1}^{n-1} \tanh\frac{\beta\epsilon_l}{2}\, 
\left[1-\prod\limits_{m=n}^\infty\tanh^2\frac{\beta\epsilon_m}{2}\right]
\end{equation}
where $s_i^0s_j^0$ is the relative orientation of $s_i$ and $s_j$ in 
the configuration with none of the surfaces excited. From this expression
we obtain the averages
\begin{equation}\label{ccf}
\overline{<s_is_j>^{2q}} \simeq \left[\prod\limits_{k=1}^{n-1} \overline{
\tanh^{2q}\frac{\beta\epsilon_k}{2}}\right]^2\,\,\overline{\left[1-\prod
\limits_{m=n}^\infty\tanh^2\frac{\beta\epsilon_m}{2}\right]^{2q}}\,\, . 
\end{equation}
At criticality, all the $\epsilon$'s in the products are independently
distributed random variables. Therefore the last factor in (\ref{ccf}) 
equals unity and we find
\begin{equation}
\overline{<s_is_j>^{2q}} \simeq e^{2n\log\overline{\tanh^{2q}\frac{\beta
\epsilon_k}{2}}}\,\, \sim L^{-q(d-2)-\eta_c^{(q)}}\,\, .
\end{equation}
Obviously, we have in general $\eta_c^{(q)}\ne q\eta_c^{(1)} \equiv q
\eta_c$. Hence, the $\eta_c^{(q)}$ are another series of independent 
critical exponents,
i.e., we have also {\em multifractal scaling of the spin-spin correlation 
function at criticality}\/. 

In \cite{S}, Sourlas considered the long-distance behaviour of correlation 
functions of critical spin glass using an effective field theory and the 
replica trick. For $d>6$, he found after averaging over the disorder that 
the different replicas are not coupled by relevant or marginal operators in 
the effective replicated Hamiltonian. It follows as in \cite{P} that $
\eta_c^{(q)}=0$ for $d>6$, i.e., the mean-field (Gaussian) behaviour. 
This is not the case in $d=6-\epsilon$ dimensions. An $\epsilon$-expansion 
around the effective field theory in $d=6$ leads to 
multifractality, i.e., $\eta_c^{(q)}\ne q\eta_c^{(1)}$, as 
found in this work by a very different approach.

\subsection{Response to a magnetic field}

\subsubsection{Introduction}

In this subsection, we will consider the magnetisation in  a small uniform 
magnetic field $H$. Many of the results obtained in this section can be 
found scattered in previous work by several authors \cite{MM,BM,BMmf,FH}. 
McMillan \cite{MM}, Bray and Moore \cite{BM,BMmf}, and Fisher and Huse 
\cite{FH,FHua} analysed the effects of a magnetic field on the {\em ordered 
phase}\/ of a spin glass by applying a variant of the argument used by Imry 
and Ma \cite{IM} to determine the stability of Ising ferromagnets to random 
fields. They find that the zero-field ordered states are destroyed beyond a 
length scale $\xi_H$ given by
\begin{equation}\label{xihT0}
\xi_H \sim \left(\frac{\Upsilon(T)}{H\sqrt{q(T)}}\right)^\frac{2}{d-2y}\,\, ,
\end{equation}
with $\Upsilon(T)\simeq J\xi_-^{-y}$ and $q(T)\simeq q_0\xi_-^\beta$. 
Equation (\ref{xihT0}) holds as long as the field $H$ is small enough so 
that $\xi_H\ga \xi_-$.  Here $\Upsilon(T)$ and $q(T)$ are the 
temperature-dependent stiffness coefficient and the Edwards-Anderson order 
parameter, respectively.
The corresponding magnetisation $m(H)$ is calculated to be
\begin{equation}\label{mhT0}
m(H) - \frac{q(T)}{y\Upsilon(T)}H \sim - \left(\frac{q^\frac{1}{2}(T)H}{
\Upsilon(T)}\right)^\frac{2y}{d-2y}\,\, .
\end{equation}
The behaviour of $m(H)$ for $T=T_c$ can then be obtained on general scaling 
grounds \cite{BMmf,MM,BM,FH}. 

The picture developed in this paper now permits us to extend the Imry-Ma 
type argument to the low-$T$ critical point. The crucial property common to 
both the droplets of the ordered phase and the composite clusters at 
criticality is that they may reverse {\em with respect to a fixed background}
\/. The effects of the magnetic field, in the ordered phase {\em and}\/ at 
criticality, can therefore be summarised as follows:
\begin{itemize}
\item The field aligns small composite clusters (thermally active or not) 
with a certain probability and larger ones with a larger probability and
\item disrupts the fixed background by aligning practically all composite 
clusters from a certain length scale ($\sim \xi_H$) on.
\end{itemize}
This allows us to present the effects of a magnetic field on the system in 
the ordered phase and at criticality in a unified way.

\subsection{A magnetic field at criticality}\label{mf}

\subsubsection{Temperatures below $T_c$}

We will first give the expressions for the correlation length $\xi_H$ in a 
field and for the magnetisation $m(H)$ for $T\laq T_c$, valid irrespective 
of the relative size of the variables $t$ and $h$. In order to do that, we 
have to know the typical reversal free energy $\Delta F^\ell$ and the 
typical magnetic energy $E^\ell_{\mbox{\tiny mag}} = H\, M^\ell$ of 
composite clusters, of linear size $\ell$, where $M^\ell$ denotes their 
total magnetisation. We recall that at length scales $\ell\ga \xi_-$ these 
composite clusters are the FH droplets. The reversal free energies $\Delta F^
\ell$ are 
\begin{equation}\label{EL}
\Delta F^\ell \sim \left\{ \begin{array}{lllcl}
J\, ,\qquad&\ell&\la&\xi_-\,\, .\\
\Upsilon(T)\ell^y\, ,\qquad&\xi_-&\la& \ell\,\, .
\end{array} \right.
\end{equation}
In the ordered phase, the mean square value of the magnetisation is $q(T)
\ell^d$ for large {\em droplets}\/. At criticality, we have to take into 
account the renormalisation of the order parameter 
\begin{equation}
q_\ell(T_c)\simeq q_0\ell^{-\frac{d-2+\eta_c}{2}}\,\, .
\end{equation}
Hence, for general $T$, the expression is
\begin{equation}\label{ML}
M^\ell \sim \left\{ \begin{array}{lllcl}
q^\frac{1}{2}_0\ell^\frac{d+2-\eta_c}{4} ,\qquad&\ell&\la&\xi_-\,\, ,\\
q^\frac{1}{2}_0\xi_-^{-\frac{\beta}{\nu}}\ell^\frac{d}{2}\, ,\qquad&\xi_-&
\la& \ell\,\, .
\end{array} \right.
\end{equation}
From simulations (see \cite{KAAHM} and references therein) and experiment 
\cite{GHD} one can infer $\eta_c\approx -0.3$. In both cases $\eta_c$ is 
obtained by hyperscaling from the values of the critical susceptibility 
exponents.

Flipping a domain of scale $\ell$ will change the system's magnetisation by 
an amount of order $\ell^\frac{d}{2}$, the magnetisation of the domain going 
from $M^\ell$ to $-M^\ell$. Half of the composite clusters can lower their 
total free free energy by aligning with the field. Thus, the system breaks 
up into domains of typically the size of a correlation volume $\xi_H^d$. The 
correlation length $\xi_H$ is obtained by comparison of the reversal free 
energy and the magnetic energy as the solution of
\begin{equation}\label{xiH}
\Delta F^{\xi_H} = E^{\xi_H}_{\mbox{\tiny mag}}\,\, .
\end{equation}
The typical fraction $f^\ell$ of domains of linear size $\ell\la \xi_H$ that 
will reverse to align with the field is 
\begin{equation}\label{fell}
f^\ell = \frac{E^\ell_{\mbox{\tiny mag}}}{\Delta F^\ell}\,\, .
\end{equation}
Finally, the total magnetisation $M(H)$ in a volume $\xi_H^d$ is given by 
the sum over all the domains, up to scale $\xi_H$, that align with the field:
\begin{equation}
M(H) = H\,\int\limits_{\sim 1}^{\xi_H}\, \frac{d\ell}{\ell}\, f^\ell\left(
\frac{\xi_H}{\ell}\right)^d\, M^\ell\,\, .
\end{equation}
Hence the magnetisation (per spin) $m(H)$ is
\begin{equation}\label{mh}
m(H) = H\, \int\limits_{\sim 1}^{\xi_H}\, \frac{d\ell}{\ell}\, \frac{(M^
\ell)^2}{\ell^d \Delta F^\ell}\,\, .
\end{equation}

Introducing (\ref{EL}) and (\ref{ML}) into (\ref{xiH}), and solving for $
\xi_H$, we get for the correlation length in a field near criticality:
\begin{equation}\label{xihres}
\xi_H = h^{-\frac{2\nu}{\Delta}}\, {\cal F}\left(\frac{t}{h^{\frac{2}{
\Delta}}}\right)\,\, ,
\end{equation}
where $\Delta = \nu(d+2-\eta_c)/2$ is the gap exponent and hyperscaling is 
tacitly understood. Numerical values for $\Delta$ can be obtained with 
the help of the relation $\Delta=\beta+\gamma$ where $\gamma$ is the 
exponent describing the divergence of the nonlinear susceptibility. 
From Monte Carlo simulations and high-temperature series expansions one has 
estimated $\beta=0.6\pm 0.1$ and $\gamma=2.95\pm 0.25$ (see \cite{KAAHM} and 
references therein). Analyses of experimental data give the same range of 
values for $\beta$ but favor a higher value of $\gamma=4.5\pm 0.3$ 
\cite{GHD}. The scaling (\ref{xihres}) at criticality has been obtained by 
Fisher and Huse \cite{FH} previously [their equation (5.13)]. The scaling 
function ${\cal F}(x)$ satisfies
\begin{equation}
\begin{array}{lcl}
{\cal F}(0)&=& \mbox{cst.}\,\, ,\\
{\cal F}(x)&{}& \mbox{analytic for $x=0$}\,\, ,\\
{\cal F}(x)&\stackrel{x\to\infty}{\sim}& (-x)^{\frac{\Delta}{(d-2y)}-1}\,\, .
\end{array}
\end{equation}
The magnetisation is obtained from (\ref{EL}), (\ref{ML}), (\ref{xiH}) and (
\ref{mh}):
\begin{equation}\label{chibTc}
m(H) - \chi H = m_{\mbox{\tiny sg}} \qquad \mbox{with $\chi\sim\frac{2}{d-2+
\eta_c}\,\frac{q_0}{J}$\,\, ,}
\end{equation}
and the singular part of the magnetisation $m_{\mbox{\tiny sg}}$ verifies
\begin{equation}\label{msg}
m_{\mbox{\tiny sg}}(t,h) \sim -\frac{2q_0^\frac{1}{2}}{d-2+\eta_c}\, h^{1+2
\frac{\beta}{\Delta}}\, {\cal M}\left(\frac{t}{h^{\frac{2}{\Delta}}}\right)\,
\, ,
\end{equation}
with a scaling function ${\cal M}(x)$ such that ${\cal M}(0)$ is a positive 
constant, and
\begin{equation}\label{msgsc}
\begin{array}{lcl}
{\cal M}(x)&{}& \mbox{analytic for $x=0$}\,\, ,\\
{\cal M}(x)&\stackrel{x\to -\infty}{\simeq}& \frac{d-2+\eta_c}{2y}\, (-x)^{
\beta-\Delta\frac{y}{d-2y}}\,\, .
\end{array}
\end{equation}
The limit $x\to -\infty$ is in agreement with the result obtained by Bray 
and Moore \cite{BM} [their equation (46)]. The linear term $\chi H$ exceeds 
the magnetisation $m(H)$ by the singular contribution $m_{\mbox{\tiny 
sg}}(t,H)$, because, just as in the ordered phase \cite{FH}, the low-field 
linear response of the domains with $L\ga\xi_H$ has saturated. 

In particular, we see from (\ref{msg}) and (\ref{msgsc}) that the exponent 
of $H$ in $m_{\mbox{\tiny sg}}$ is $1+2\frac{y}{d-2y}$ and $1+2\frac{\beta}{
\Delta}$ for $x\to 0$ and $x\to -\infty$, respectively. Hence, the singular 
contribution is dominated by the linear response. However, both values of 
the exponent are smaller than $3$, so that the singular term dominates over 
the $H^3$ term in the magnetisation and causes the zero-field nonlinear 
susceptibility to be infinite. Furthermore, we see from (\ref{msg}) that the 
third derivative of the magnetisation diverges, for $H \to 0$, {\em at} the 
critical point as
\begin{equation}
-\frac{d^3m}{dH^3} \simeq \frac{2q_0^\frac{1}{2}}{d-2+\eta_c}\,\left(
\frac{q_0^\frac{1}{2}}{J}\right)^{1+2\frac{\beta}{\Delta}}\, H^{2(\frac{
\beta}{\Delta}-1)}\,\, .
\end{equation}
For dimensions $d$ close above $d_\ell$ the following inequality holds:
\begin{equation}\label{inequal}
\frac{y}{d-2y} < \frac{\beta}{\Delta}\,\, \Leftrightarrow\,\, d < \beta
\frac{d-y}{\nu y}\,\, .
\end{equation}
The second one follows from the fact that, for $d\searrow d_\ell$, we have $
\nu y\to \frac{1}{2}$ \cite{FH}, $y\searrow 0$ and $\beta \sim y^{-
\frac{1}{2}}\nearrow \infty$ (see \cite{FH}). It seems reasonable to assume 
that (\ref{inequal}) is valid in general dimension $d>d_{\ell}$, so that the 
divergence for $H\to 0$ of $-\frac{d^3m}{dH^3}$ is stronger in the ordered 
phase than at criticality. 

Summarising the above results for $H\to 0$ at $t$ fixed, we get:
\begin{equation}
\begin{array}{llcl}
T<T_c:\qquad&m(H)&=& a_{10}H-a_s(t)H^{1+\frac{2y}{d-2y}} + \ldots\\[1mm]
{}&{}&{}&\,\, a_s(t) \simeq a_{s0}\,|t|^{\beta-\frac{\Delta y}{d-2y}}\\[2mm]
T=T_c:\qquad&m(H)&=& a_{10}H-a^c_sH^{1+\frac{2\beta}{\Delta}} + \ldots
\end{array}
\end{equation}
Here, $a_{10}$, $a_{s0}$, and $a^c_s$ are amplitudes, and (\ref{inequal}) 
implies that the exponent of $a_s(t)$ is positive.

\subsubsection{Temperatures above $T_c$}

Let $\xi_+$ be the zero-field correlation length for $T>T_c$. Domains of 
linear size $\ell\la \xi_+$ behave critically, whereas much larger domains 
are paramagnetic. When $H$ is small, domains of size up to $\xi_+^d$ align 
partially and in order to obtain the magnetisation we have to integrate 
until scale $\xi_+$. When $H$ is larger, there is $\xi_H < \xi_+$, and the 
integration runs until $\xi_H$ as in equation (\ref{mh}). In the latter 
case, we get again the result (\ref{msg}), but with small positive argument 
of the scaling function. In the former case, we have 
\begin{equation}
m(H) = H\,\int\limits_{\sim 1}^{\xi_+}\, \frac{d\ell}{\ell}\frac{(M^\ell)^2}{
\ell^d\Delta F^\ell} \,\, ,
\end{equation}
so that
\begin{equation}\label{msg+}
m(H) - \chi H \sim - \frac{2}{d-2+\eta_c}\,\frac{q_0}{J}\, t^\beta\, H\,\, ,
\qquad \mbox{with}\quad \chi=\frac{2}{d-2+\eta_c}\,\frac{q_0}{J}\,\, ,
\end{equation}
the singular part of the magnetisation contributing a term $\sim t^\beta$ to 
the linear susceptibility.

We can write the magnetisation for $T\gaq T_c$, from (\ref{msg}) and (
\ref{msg+}), in a compact form:
\begin{equation}
m(H) - \chi H \sim m_{\mbox{\tiny sg}}\,\, ,
\end{equation}
with a constant contribution of the nonsingular part of the magnetisation to 
the linear susceptibility as in (\ref{msg+}), whereas the singular part of 
the magnetisation behaves as
\begin{equation}\label{msgres}
m_{\mbox{\tiny sg}}(t,h) = -\frac{2q_0^\frac{1}{2}}{d-2+\eta_c}\, h^{1+2
\frac{\beta}{\Delta}}\, {\cal M}\left(\frac{t}{h^\frac{2}{\Delta}}\right)\,
\, ,
\end{equation}
with ${\cal M}(0)$ a positive constant and the scaling function ${\cal 
M}(x)$ satisfies
\begin{equation}\label{msgressc}
\begin{array}{lcl}
\mbox{${\cal M}(x)$}&{}&\mbox{analytic for $x=0$}\,\, ,\\
{\cal M}(x)&\stackrel{x\to \infty}{\simeq}&  x^{\beta}\,\, .
\end{array}
\end{equation}

To study the limit $h\to 0$ at small positive $t$ of (\ref{msgressc}), we 
write first
\begin{equation}
\begin{array}{lcl}
m_{\mbox{\tiny sg}}(t,h) &\sim& h^{1+2\frac{\beta}{\Delta}}\, {\cal M}\left(
\frac{t}{h^\frac{2}{\Delta}}\right)\\
{}&=&h^3\,t^{-\gamma}\hat{\cal M}\left(\frac{h^2}{t^\Delta}\right)\,\, ,
\end{array}
\end{equation}
where $\hat{\cal M}\left(\frac{h^2}{t^\Delta}\right)\equiv \frac{t^
\Delta}{h^2}{\cal M}\left(\frac{t}{h^\frac{2}{\Delta}}\right)$. From (
\ref{msgressc}), we have
\begin{equation}
\hat{\cal M}(u)\simeq u^{-1}\,\, ,\qquad u \to 0\,\, .
\end{equation}
For $t>0$, the function $m_{\mbox{\tiny sg}}$ should actually be {\em 
nonsingular}\/ in $h$. Hence, with the assumption that $\hat{\cal M}$ is 
analytic in $h^2$, we get in summary for the magnetisation above the 
critical point:
\begin{equation}\label{resabTc}
\begin{array}{llcl}
T>T_c:\qquad&m(H)&=& a_1(t)H+a_3(t)H^3 +\, \ldots\\[1mm]
{}&{}&{}&\,\, a_1(t) \simeq a_{10}-a_{1\beta}t^\beta\,\, ,\\
{}&{}&{}&\,\, a_3(t) \simeq a_{1\gamma}t^{-\gamma}\,\, ,
\end{array}
\end{equation}
where $a_{10}$, $a_{1\beta}$, and $a_{1\gamma}$ are amplitudes. This shows 
that, for $t\searrow 0$, the linear susceptibility $\chi$ has a nondiverging 
singular contribution $\sim t^\beta$ and the well-known feature \cite{BMmf} 
that the nonlinear susceptibility diverges as $t^{-\gamma}$. 

\subsubsection{A magnetic field at $d=d_\ell$}

We now turn to the special case of applying a small magnetic field $H$ {\em 
at} the lower critical dimension $d=d_\ell$. For $d\searrow d_\ell$, the 
contribution of the nonsingular part of the magnetisation to the linear 
susceptibility [see (\ref{chibTc})] diverges, because $\eta_c \nearrow 2-d$. 
However, due to the simultaneous divergence of the singular part of the 
magnetisation [see (\ref{msg}), (\ref{msgres})] the linear susceptibility 
remains finite for $T>0$, because the leading contributions of the 
nonsingular and of the singular part cancel each other in the limit $d=d_
\ell$, as previously observed by \cite{MM,BMmf}. This gives rise to a 
logarithmic singularity of the magnetisation, for $H \to 0$ at $T=0$, which 
can be seen as follows. To obtain the expression of the magnetisation at 
$d=d_\ell$, we can reason as in the preceding subsection for $T\gaq T_c$, 
with here $T_c(d_\ell)=0$. At $T=0$, the reversal energy $\Delta F^\ell$ and 
the magnetisation $M^\ell$ of a cluster of linear size $\ell$ are
\begin{equation}
\begin{array}{lcl}
\Delta F^\ell&\sim& J\,\, ,\\
M^\ell&\sim& q_0^\frac{1}{2}\ell^\frac{d_\ell}{2}\,\, ,
\end{array}
\end{equation}
for all length scales $\ell$, so that from (\ref{xiH})
\begin{equation}
\xi_H = \left(\frac{J}{q_0^\frac{1}{2}H}\right)^\frac{2}{d_\ell}\,\, .
\end{equation}
However, for $T>0$, the renormalisation of the reversal free energy $\Delta 
F^\ell$ is nontrivial. Solving (\ref{KlogL}) for $y=0$ we obtain 
\begin{equation}
\Delta F^\ell\sim J\, \sqrt{1-c\left(\frac{k_{\mbox{\tiny B}}T}{J}\right)^2
\log\ell}\,\, ,
\end{equation}
as McMillan did previously \cite{MM}. The magnetisation, for $H\to 0$ at $T
\gaq 0$, is readily calculated in the same way as above for $d> d_\ell$ and 
$T\gaq T_c$, the zero-field correlation length being $\xi_+ \simeq e^{-
\frac{1}{2c{\mit f_+}^2}+\frac{J^2}{2c(k_{\mbox{\tiny B}}T)^2}}$\, [see (
\ref{dlcl})]. One finds
\begin{equation}
m(T,h) = \frac{q_0^\frac{1}{2}J}{k_{\mbox{\tiny B}}}\,\frac{h}{T}\, {\cal 
M}_{d_\ell}\left(k_{\mbox{\tiny B}}T\log h^{-1}\right)\,\, ,
\end{equation}
with a scaling function ${\cal M}_{d_\ell}$ that satisfies
\begin{equation}
\begin{array}{lcl}
{\cal M}_{d_\ell}(x)&\stackrel{x\to 0}{\simeq}& x\,\, ,\\
{\cal M}_{d_\ell}(x)&\stackrel{x\to \infty}{\simeq}&  1\,\, .
\end{array}
\end{equation}
Thus, in the limit $H\to 0$ at $T$ fixed, we obtain the Curie law as 
expected in the paramagnetic phase on general theoretical grounds 
\cite{vEvH}.

\setcounter{equation}{0}
\section{Some comments on critical dynamics}\label{dyn}

Random systems, like random field
magnets and spin glasses, have far slower dynamics than conventional
pure systems (see e.g.~\cite{FHF,rev1,rev2} and references therein). This is 
often attributed to fact that the competition between various types of 
interactions leads to free energy barriers of height $B^L$ that have to be 
crossed in thermally activated dynamical processes. If a static theory like
the one of the preceding sections is to be extended to include dynamical
phenomena, then {\em additional}\/ assumptions concerning these barriers
have to be made. It is generally assumed that a barrier typically grows
with the length scale $L$ involved in the process as 
\begin{equation}\label{BL}
B^L \sim L^\psi\,\, ,
\end{equation}
with a {\em barrier exponent}\/ $\psi>0$ and gives naturally rise to long 
time scales 
\begin{equation}
\tau^L\sim \exp(\frac{B^L}{k_{\mbox{\tiny B}}T})\,\, .
\end{equation}
This type of scaling is called {\em activated dynamic scaling}. 
The activated processes can be either fluctuations in equilibrium 
or involve approach to equilibrium.

Generally, it has been agreed upon that this is indeed the mechanism valid 
at sufficiently large length scale in the spin glass ordered phase. However, 
it has remained unclear whether the {\em critical}\/ dynamics in these 
systems should be described by activated or rather by conventional 
(power-law) dynamic scaling. Until now, most often critical dynamics has 
been scaled in a conventional manner, with anomalously large exponents, 
although thermally activated critical dynamic scaling with a small barrier 
exponent also appears to account reasonably well for experimental spin-glass 
behaviour at the critical point in three dimensions and does not seem to be 
incompatible with numerical simulations \cite{rev2,H}. Again, however, the 
scaling with either approach can be corrupted in $d=3$ due to a number of 
serious problems (see \cite{GHD,KAAHM} and references therein for details). 

The picture of this paper represents an extension of the droplet picture of 
the spin glass into the critical region. We argue in the following that it 
suggests that spin glass dynamics is activated also {\em at the critical 
point}\/, at least in dimensions close above $d_\ell$. We recall that in the 
FH theory the droplet walls minimise the free energy of a reversed spin 
domain. They are locally optimal and the fact that they dominate the
physics of the ordered phase reflects the strong spatial inhomogeneity of 
thermal fluctuations in that phase. In particular, to reverse a droplet of 
scale $L$, the domain wall must pass through regions of high free-energy 
cost which are larger than the excitation free energy of the completed 
droplet. Therefore FH argue that droplet reversal is a thermally activated 
process with barriers $B^L$ behaving as in (\ref{BL}) and growing at least 
as fast with length scale as the free energy difference between initial and 
reversed state. Therefore its barrier exponent satisfies
\begin{equation}\label{ypsi}
y(d)\laq \psi(d)\,\, 
\end{equation}
in dimensions $d$ above $d_\ell$ \cite{FH}. 

The same reasoning can be applied to the protodroplets of low-dimensional 
spin glass at criticality. But the lower bound $0\laq \psi_c$ for the 
scaling of the barriers at criticality, obtained from comparison with the 
scaling of the free energy, is trivial. Indeed, $\psi_c=0$ would mean 
conventional scaling. 
We now give an argument that points towards $\psi_c(d)>0$ in an interval
of dimensions extending upward from $d_\ell$. Dynamics near zero 
temperature in two 
dimensions has been argued on the basis of numerical simulations and  
experiments to be thermally activated. Numerical work \cite{psi2} points to $
\psi(2)\approx 1$ (but see also \cite{rev2} and references therein) and 
experiments give $\psi(2)/|y(2)| = 1.6 \pm 0.2$ and $\psi(2)/|y(2)| = 
0.9\pm 0.1$ for $40${\AA} \cite{psi2exp} and $\la 20${\AA} \cite{Matt} 
thick spin glass films respectively. This would mean $\psi(d)>0$, and, 
since the barrier exponent is expected to increase with dimension, 
also $\psi(d_\ell)>0$. Hence by continuity $\psi_c(d)>0$ in an interval 
of dimensions upward from $d_\ell$. 
In these dimensions the barrier heights scale at criticality like
\begin{equation}
B^L(T_c) \sim J\, L^{\psi_c}\,\, ,\qquad 0<\psi_c\,\, .
\end{equation}
Hence, the characteristic reversal time $\tau^L$ of a 
composite cluster at a large scale $L$ at criticality  
is exponentially length scale dependent:
\begin{equation}
\tau^L\sim \tau_0\exp\frac{JL^{\psi_c}}{k_{\mbox{\tiny B}}T_c}\,\, .
\end{equation}
Such a behaviour implies \cite{FH} that ac correlation functions should 
scale, in the critical region, as functions of $\frac{\log(\omega\tau_0)}{
\xi_-^{\psi_c}}$ and that  
\begin{equation}
\chi_3(3\omega;\omega)\sim |\log^\frac{\gamma}{\psi_c\nu}(\omega\tau_0)|\,\, 
,
\end{equation}
i.e., the cubic nonlinear susceptibility diverges only logarithmically with 
vanishing frequency at criticality.

\setcounter{equation}{0}
\section{Conclusion}\label{concl}

We have analysed the critical state of the low-dimensional Ising spin glass,
in particular in the limit of the dimension $d$ approaching the lower
critical dimensionality $d_\ell$.  

The best existing approximate theory for the low-dimensional spin glass
is the scaling or droplet theory developed 
since the middle 1980's by McMillan, Bray and Moore, Fisher and Huse,
and others \cite{MM,BM,FH,FHua}.
The building blocks of this theory, in the language of Fisher and Huse,
are the low free energy excitations, called droplets, from the spin
glass ordered state. Droplets are reversed regions of one phase embedded
in the opposite phase. Their excitation energy has a random value on an
energy scale that increases with their size. 
The droplet wall, like in ferromagnetism, has a width of the order of the
correlation length. Consequently droplets exist only in the
ordered phase and the droplet theory in its original formulation
says nothing about criticality.

We show in this work that the FH theory has a natural extension into the
critical region, that is, to length scales less than the correlation
length. This extension can be presented as a heuristic expansion
around the dimension $d_\ell$. Its main idea is summarised as follows.

In a critical state very close to $d_\ell$ most
bonds are frozen, but there is a very dilute set of surfaces
(of width zero) that enclose clusters of spins able to thermally reverse
with respect to their environment. The density $p$ of these surfaces is
the expansion parameter; it is proportional to the critical temperature 
$T_c$.
Because of scale invariance there is the same density $p$ of thermally active
surfaces at all length scales. Since $p$ is small but nonzero,
interactions between these surfaces (through common surface sections)
should be taken into account.
Therefore, {\em in the low-dimensional critical state the thermally active
degrees of freedom are organised in finite disjoint sets of interacting
thermally active surfaces}\/. These sets have a distribution of sizes that
ranges to infinity. Each set has a self-similar structure on scales from
its own size down to the lattice cutoff. Such a set of interacting
surfaces is appropriately called a {\em protodroplet}\/. In the off-critical
limit, that is, for length scales beyond $\xi_-$,
one of the surfaces participating in a protodroplet will dominate in
size so that all the others may be considered as thermally decorating this
large one and building up its wall: the protodroplet has then become a
droplet.  

This picture of the critical state explains the phenomenon of critical
chaos \cite{NH1,NH2}. It has enabled us to discuss the effects of a magnetic 
field in a unified way, using an Imry-Ma type argument in the ordered phase {
\em and}\/ at criticality. Finally, we have briefly argued, on the basis of 
additional hypotheses, that
in low enough dimensions the critical dynamics is thermally activated.

The picture of this paper has the practical advantage of making definite 
statements about spin glass properties on length scales {\em below}\/ the
correlation length (in contrast to the Fisher and Huse droplet picture,
which is valid on scales {\em above}\/ the correlation length). Some of
our predictions should be easily accessible in  
simulations of the Edwards-Anderson model. One may of course attempt to
determine the existence and properties of the thermally active surfaces
(they exist from the scale of the lattice distance up), but also more
conventionally verify our predictions for various spatial correlation
functions. 

The following three remarks about possible future
sinulations  are in no way meant to be exhaustive.
\begin{enumerate}
\item {\em About thermally active cluster surfaces.}\/ A quantity that
measures the degree of thermal activity of the surface element
perpendicular to a bond between two neighbouring spins $s_i$ and $s_{i'}$ is
$|<s_i s_{i'}>|$. If this quantity is zero, the bond is strongly
thermally active; if it is close to unity the bond is nearly frozen.
Its spatial (or disorder) average is equal to $1-{\cal O}(k_BT_c/J)$.
Conceivably a chart of all nearest neighbour $|<s_i s_{i'}>|$ could be
converted at the end of the simulation into a map of all thermally
active surfaces. 
\item {\em Dimension dependence.}\/ As the dimension $d$ moves up 
from $d_\ell$, the critical temperature $T_c$ moves away from $0$ and 
thermally active cluster surfaces 
become gradually harder to distinguish, since almost all bonds will
acquire some degree of thermal activity. Also, the thermally active
surfaces, even if they 
can be distinguished, may, from some dimension on, form infinite percolating 
structures and the
theory of our paper is not directly applicable.
In dimension $d=2$ and at sufficiently low temperature, the thermally
active cluster surfaces exist on length scales less than the correlation
length $\xi$, but the correlation functions are not multifractal; beyond
$\xi$ the surfaces form an infinitely connected structure. 
Therefore d=3 is the ideal dimension for testing our predictions.
\item {\em Analysis in terms of correlation functions.}\/
This is the more conventional approach. The theory predicts power laws
at distances less than the correlation length (but of course
sufficiently large with respect 
to the lattice cutoff; how much larger we do not know). The powers are
in general different for different moments of a given correlation
function. We extract from the preceding sections what seem to us the
most important ones. 

The disorder
averaged energy-energy correlation function $\overline{<\Delta e_i \Delta 
e_j>}$, equation (\ref{defencorr}), should behave as (see equations (
\ref{defg2}) and (\ref{G2r})):
\begin{equation}
\overline{<\Delta e_i \Delta e_j>}  \sim  |i-j|^{2y_c-2d}  \quad\qquad  (d>d_
\ell\, ,\quad |i-j| < \xi_-)\,\, .
\end{equation}
However, the theory also says that $<\Delta e_i \Delta e_j>$ has 
fluctuations larger than its average value. Its variance (see
equations (\ref{fc}) and (\ref{G4r})) is
\begin{equation}
\overline{<\Delta e_i \Delta e_j>^2} - \overline{<\Delta e_i \Delta e_j>}^2  
\sim |i-j|^{4\zeta_c-2d}\,\, .
\end{equation}
(The Migdal-Kadanoff values for $d=3$ are $y_c\approx 0.36$ and 
$\zeta_c\approx 0.57$ \cite{NH1,NH2,TH}.) We are not aware of 
any direct determination yet 
of the exponent $\zeta_c$ in the Edwards-Anderson model. 
The corresponding zero temperature exponent $\zeta$ has in $d=3$ the 
Migdal-Kadanoff value $\zeta\approx 0.75$ \cite{NH1,NH2,TH}; the difference $
\zeta-\zeta_c$,  just as the difference $y_c-y$, measures the distance to 
the lower critical
dimension. The higher moments $\overline{<\Delta e_i \Delta e_j>^{2q}}$ and  
$\overline{<\Delta e_i \Delta e_j>^{2q-1}}$ decay with
powers of $|i-j|$ that involve a hierarchy of exponents $\zeta_c(q)$, see
equation (\ref{G2rm}).  
  
A quantity that should be measurable with relatively good precision is
the fourth energy cumulant $\overline{<\Delta E^4>}_{\mbox{\tiny cum}}$ 
(equation (\ref{4thcres})) in an arbitrary subvolume of linear dimension 
less than the correlation length. As $V$ becomes large, 
\begin{equation}
\frac{\overline{<\Delta E^4>}_{\mbox{\tiny cum}}}{V} 
\stackrel{V\to \infty}{\sim} \left\{ \begin{array}{r@{\qquad\mbox{if}
\quad}lr}
V^{4\frac{\zeta_c}{d}-1} & 4\zeta_c/d > 1\,\, . \qquad&\mbox{(a)} \\
  \mbox{cst. $-$ (terms nonanalytic in $V$)} & 4\zeta_c/d < 1\,\, . \qquad&
\mbox{(b)} 
\end{array} \right.
\end{equation}
The MKRG values indicate that in $d=3$ the second scenario
will be realised, since numerically $4\zeta_c/d-1 = 4\cdot 0.57/3-1 = -0.24$.
However, we cannot be sure of the sign of this quantity in $d=3$. 
At $d_\ell$, MKRG gives for the same quantity the value
of approximately $4\cdot 0.75/2.5-1 = 0.2$. Hence the fourth energy
cumulant in $d=3$ provides a - possibly rather sensitive - way of determining
$\zeta_c$. 
\end{enumerate}

\vspace{2cm}

\noindent {\bf Acknowledgements.} One of us (M.J.T.) wishes to thank David 
Huse
for many valuable discussions related to the subject.

\end{document}